\documentclass[useAMS,usenatbib,onecolumn]{mn2e}

\usepackage[pdftex]{graphicx}

\title[On the photoelectric quantum yield of small dust particles]{On the photoelectric quantum yield of small dust particles}
\author[Hiroshi Kimura]{Hiroshi Kimura\thanks{E-mail:
hiroshi\_kimura@cps-jp.org}\thanks{CPS Collaborating Scientist.}\thanks{Current address: CPS (Center for Planetary Science), 7-1-48 Minatojima Minamimachi, Chuo-ku, Kobe 650-0047, Japan}\\
Graduate School of Science, Kobe University, 1-1 Rokkodai-cho, Nada-ku, Kobe 657-8501, Japan}
\begin{document}

\date{Accepted 2016 April 07. Received 2016 April 05; in original form 2015 February 13}

\pagerange{\pageref{firstpage}--\pageref{lastpage}} \pubyear{2016}

\maketitle

\label{firstpage}

\begin{abstract}
Photoelectron emission is crucial to electric charging of dust particles around main-sequence stars and gas heating in various dusty environments.
An estimate of the photoelectric processes contains an ill-defined parameter called the photoelectric quantum yield, which is the total number of electrons ejected from a dust particle per absorbed photon.
Here we revisit the so-called small particle effect of photoelectron emission and provide an analytical model to estimate photoelectric quantum yields of small dust particles in sizes down to nanometers.
We show that the small particle effect elevates the photoelectric quantum yields of nanoparticles up to by a factor of $10^3$ for carbon, water ice, and an organics, and a factor of $10^2$ for silicate, silicon carbide, and iron.
We conclude the surface curvature of the particles is a quantity of great importance to the small particle effect, unless the particles are submicrometers in radius or larger.
\end{abstract}

\begin{keywords}
dust, extinction --- interplanetary medium --- meteorites, meteors, meteoroids --- zodiacal dust).
\end{keywords}

\section{Introduction}

It is inevitable that dust particles carry electric charges on their surfaces as a consequence of photoelectron emission, electron and ion attachments, and secondary electron emission.
There is a consensus that photoelectron emission is the dominant charging process for dust particles in the vicinity of a main-sequence star \citep{belton1966,mukai1981,horanyi1996,kimura-mann1998}.
Photoelectron emission is also an important process to heat gas in the interstellar medium, planetary nebulae, the intergalactic medium, and disks around young stellar objects \citep{draine1978,dopita-sutherland2000,inoue-kamaya2010,pedersen-gomezdecastro2011}.
An estimate of the photoelectric current requires knowledge of the photoelectric quantum yield that is defined as the number of photoelectrons ejected from a surface per absorbed photon.
The dependence of photoelectric yields on photon energy has been determined by laboratory experiments on photoelectron emission \citep[e.g.,][]{berglund-spicer1964b}.
Although commonly used materials in the experiments are irrelevant to cosmic dust environments, laboratory experiments with astrophysically interesting materials are available for graphite, silicon carbide, lunar surface material, silica with a carbon coating, and vitreous carbon \citep{taft-apker1955,philipp1958,feuerbacher-et-al1972,feuerbacher-fitton1972,willis-et-al1973a,willis-et-al1973b}.

Since laboratory experiments on photoelectron emission are commonly performed with slab surfaces, photoelectric quantum yields of slab surfaces are often substituted for those of dust particles.
However, if the size of the particles goes down below submicrometers, then the difference in photoelectron emission between dust particles and slab surfaces becomes significant and needs to be addressed properly.
On the basis of Mie theory, \citet{watson1973} predicted enhanced photoelectric yields for small spherical particles in comparison to semi-infinite slab surfaces.
\citet{ballester-et-al1995} applied \citeauthor{watson1973}'s model to calculate the photoelectric quantum yields of graphite and silicate grains for the purpose of modeling photoelectron emission from interstellar dust.
\citet{draine1978} provided an analytic formula that reproduces the enhancement of the photoelectric quantum yield given by \citeauthor{watson1973}'s model with an accuracy of about 20\%.
Because of its simplicity, \citeauthor{draine1978}'s formula has been commonly used to estimate the enhancement of the photoelectric quantum yield for small dust particles \citep[e.g.,][]{bakes-tielens1994,weingartner-draine2001,weingartner-et-al2006}.
However, \citet{watson1973} himself noted that the model underestimates the photoelectric quantum yields of small particles, because the curvature of the particle surface was not taken into account.
The curvature of the surface is also known to elevate the work function, which is the threshold of prohibiting low-energy electrons to leave the surface \citep{mueller-et-al1991}.
As a matter of fact, \citet{watson1973} did not take into account the work function, thus implicitly assumed that no electrons are subject to the work function.

The so-called small particle effect on the photoelectric yield has been observed experimentally for very fine metallic nanoparticles \citep{schmidtott-et-al1980,mueller-et-al1988a,schleicher-et-al1993}.
However, a more recent series of experiments on photoelectron emission from submicrometer-sized grains have revealed contradictory results \citep{abbas-et-al2002,abbas-et-al2006,abbas-et-al2007}.
The small particle effect should play a vital role in the dynamics of electrically charged nanoparticles, provided that they are exposed to ultraviolet stellar radiation.
Such a case is apparently true of the nanoparticles detected in the plumes of Jupiter's moon Io and Saturn's moon Enceladus as well as in the solar wind \citep[see,][]{graps-et-al2000,hill-et-al2012,meyervernet-et-al2009}.
In addition to the experimental demonstration of the small particle effect, the recent in-situ detections of nanoparticles in the Solar System persuades us to revisit \citeauthor{watson1973}'s model on the photoelectric quantum yield of small dust particles.

\section[]{Model}

\subsection{Photoelectric quantum yield of small dust particles}

The photoelectric quantum yield is defined by the number of electrons ejected from a particle per absorbed photon.
The photoelectron yields for spherical particles have been studied by several authors based on the well-know three-step model but with different strategies \citep{watson1973,penn-rendell1981,penn-rendell1982,mueller-et-al1991}.
In the three-step model, an electron (1) is excited by absorption of a photon, (2) reaches the surface without suffering an inelastic collision, and (3) escapes from the surface \citep{smith1971}\footnote{Note that the emission of Auger electrons and secondary electrons induced by high-energy ($h\nu > 100~\mathrm{eV}$) photons may play an important role in the determination of the photoelectric quantum yields in quasar host galaxies, supernova remnants, and the central regions of galaxy clusters \citep{weingartner-et-al2006}. However, we do not consider such emission in this study, because the contribution of low-energy photons dominates the photoelectric current in the Solar System \citep{feuerbacher-et-al1972,feuerbacher-fitton1972,willis-et-al1973a}.}.
We extend their studies to take into account the surface curvature and potential barrier $W$ for the escape probability of electrons and the size dependence of work function.
The photoelectric quantum yield ${Y}_{a}(h\nu)$ in the three step model can be described by\footnote{While a geometrical optics approach was used by \citet{dwek-smith1996}, we adopt a Mie scattering approach, since photon energies considered in this paper are not high enough to the former approach.}
\begin{eqnarray}
{Y}_{a}(h\nu) = \frac{\int dV \, {p}_{\mathrm{esc}}(r) \, {\bf E}^{\ast}(r, \theta, h\nu) \cdot {\bf E}(r, \theta, h\nu)}{\int dV \, {\bf E}^{\ast}(r, \theta, h\nu) \cdot {\bf E}(r, \theta, h\nu)} ,
\label{pee-sphere}
\end{eqnarray}
where ${p}_{\mathrm{esc}}(r)$ is the probability to escape from the surface for an electron generated at a distance $r$ from the center of the spherical particle and ${\bf E}(r, \theta, h\nu)$ is the electric field vector inside the element of volume $dV = drd\Omega$.
The integration of ${\bf E}^{\ast} \cdot {\bf E}$ over solid angle $\Omega$ can be performed analytically \citep{watson1973};
\begin{eqnarray}
\int_{}^{} {\bf E}^{\ast} \cdot {\bf E} \, d\Omega 
= \frac{2\pi}{{k}^{2}{r}^{2}} \sum_{n=1}^{\infty}\left({2n+1}\right)\left\{{{\left|{{c}_{n}}\right|}^{2} \left[{1 + {\left|{{D}_{n}({mkr})}\right|}^{2} - \frac{1}{{\left|{m}\right|}^{2}} \left({1 - \frac{n(n+1)}{{k}^{2}{r}^{2}}}\right)}\right] + {\left|{{d}_{n}}\right|}^{2}}\right\} {\left|{{\Psi}_{n}({mkr})}\right|}^{2} ,
\end{eqnarray}
where ${c}_{n}$ and ${d}_{n}$ are the Mie scattering coefficients of the field inside the particle, $m(h \nu)$ is the complex refractive index at photon energy $h \nu$, ${\Psi}_{n}\left({mkr}\right)$ is the Riccati-Bessel function derived from the Bessel function of first kind, and ${D}_{n}\left({mkr}\right)$ is the logarithmic derivative of the Riccati-Bessel function \citep{bohren-huffman1983}.

Taking into account the surface curvature of a spherical particle and the potential barrier at the surface, the escape probability for the electrons generated isotropically is given by
\begin{eqnarray}
{p}_{\mathrm{esc}}(r) = \frac{1}{2} \int_{0}^{\pi } \exp \left({-\frac{l}{{l}_{\mathrm{e}}}}\right) \sin \varphi \, d\varphi - \frac{1}{2} \int_{{\varphi }_{\mathrm{c}}}^{\pi - {\varphi }_{\mathrm{c}}} \exp \left({-\frac{l}{{l}_{\mathrm{e}}}}\right)\sin \varphi \, d\varphi,
\label{p_escape}
\end{eqnarray}
where ${l}_{\mathrm{e}}$ is the mean free path for inelastic-scattering characteristic of electrons with energy $E$.
The second term of Eq.~(\ref{p_escape}) implies that produced electrons can escape from the surface only if a component of their kinetic energy perpendicular to the surface exceeds the work function $W$ \citep{fowler1931,houston1937,smith1971,penn-rendell1982}.
The distance $l$ for the electrons to reach the surface in the direction having angle $\varphi$ from the radial direction is given by
\begin{eqnarray}
l = \sqrt{{a}^{2}- {\left({r\sin {\varphi }}\right)}^{2}}-r\cos \varphi ,
\label{l}
\end{eqnarray}
If $r > a\sqrt{1-W/E}$, the critical angle ${\varphi }_{\mathrm{c}}$ is given by
\begin{eqnarray}
\cos {\varphi }_{\mathrm{c}} = \sqrt{1 - {\left({\frac{a}{r}}\right)}^2 \left({1 - \frac{W}{E}}\right)} ,
\label{critical_angle}
\end{eqnarray}
while for $r \le a\sqrt{1-W/E}$
\begin{eqnarray}
\cos {\varphi }_{\mathrm{c}} = 0 .
\end{eqnarray}

We use an empirical formula for the dependence of the mean free path ${l}_{\mathrm{e}}$ on electron energy $E$ given by:
\begin{eqnarray}
{l}_{\mathrm{e}}(E) = a \left({\frac{E}{1~\mathrm{eV}}}\right)^{-2}+b \left({\frac{E}{1~\mathrm{eV}}}\right)^{1/2},
\label{l_e}
\end{eqnarray}
where $a$ and $b$ are material-dependent parameters \citep{seah-dench1979}.
Since the model does not consider the energy distribution of electrons at their production in a particle, we shall hereafter take $E = h\nu$. 

\citet{schmidtott-et-al1980} observed an increase in the work function for small silver spherical particles with decreasing the radius of the particles.
On the basis of theoretical investigation, \citet{smith1965} and later \citet{wood1981} found that the work function is greater for small spheres than for large planar surface.
The work function for a sphere is estimated by \citep{brus1983,makov-et-al1988,wong-et-al2003}
\begin{eqnarray}
W = {W}_{\infty} +\frac{1}{4\pi {\epsilon}_{0}}\frac{3}{8}
\frac{{e}^{2}}{a}\frac{\epsilon - 1}{\epsilon} ,
\label{work-function}
\end{eqnarray}
where ${W}_{\infty}$ is the work function for a bulk, ${\epsilon}_{0}$ is the permittivity of free space, and $\epsilon$ is the static dielectric constant of the grain material relative to vacuum.

\subsection{Photoelectric quantum yield of a semi-infinite slab \label{slab}}

In order to elucidate the small-particle effect, it is a common practice to compare the photoelectric quantum yields between a small dust particle and a semi-infinite slab of the same composition.
The three-step model is equally applicable to photoelectron emission from a semi-infinite slab irradiated by UV rays at normal incidence to the surface.
The probability of absorption at $x$ from the entry into the slab having a complex refractive index $m$ is given by
\begin{eqnarray}
\frac{1}{{l}_{\mathrm{a}}}\exp\left({-\frac{x}{{l}_{\mathrm{a}}}}\right) \, dx ,
\label{p_abs_bulk}
\end{eqnarray}
where ${l}_{\mathrm{a}}$ is the photon attenuation length, which is given by
\begin{eqnarray}
{l}_{\mathrm{a}}(h \nu)=\frac{c}{|4\pi \nu \Im{(m)}|},
\end{eqnarray}
with $c$ being the speed of light in vacuum.
The escape probability for photoelectrons is given by
\begin{eqnarray}
\frac{1}{2} \exp\left({-\frac{l}{{l}_{\mathrm{e}}}}\right)\sin \varphi \, d\varphi ,
\label{p_esc_bulk}
\end{eqnarray}
where $l={x}/{\cos \varphi} $ is the distance to the surface of a semi-infinite slab from the point where the photoelectrons are generated.
Eqs.~(\ref{p_abs_bulk}) and (\ref{p_esc_bulk}) yield \citep[cf.][]{berglund-spicer1964a}
\begin{eqnarray}
{Y }_{\infty}\left(h\nu \right) = \int_{0}^{\infty}\frac{1}{{l}_{\mathrm{a}}}\exp\left({-\frac{x}{{l}_{\mathrm{a}}}}\right) \, dx \int_{0}^{{\varphi }_{\mathrm{c}}}\frac{1}{2} \exp\left({-\frac{l}{{l}_{\mathrm{e}}}}\right)\sin \varphi \, d\varphi,
\label{pee-bulk0}
\end{eqnarray}
where the critical angle ${\varphi }_{\mathrm{c}}$ is given by
\begin{eqnarray}
\cos {\varphi }_{\mathrm{c}} = \sqrt{\frac{W_{\infty}}{E}} ,
\label{critical_angle_bulk}
\end{eqnarray}
for an electron with energy $E$.
By integrating Eq.~(\ref{pee-bulk0}) analytically, we obtain
\begin{eqnarray}
{Y }_{\infty}\left(h\nu \right) = \frac{1}{2}\left[{1-\sqrt{\frac{{W}_{\infty}}{h\nu}}+\frac{{l}_{\mathrm{a}}}{{l}_{\mathrm{e}}}\log \left({\frac{\sqrt{\frac{{W}_{\infty}}{h\nu}}+\frac{{l}_{\mathrm{a}}}{{l}_{\mathrm{e}}}}{1+\frac{{l}_{\mathrm{a}}}{{l}_{\mathrm{e}}}}}\right)}\right] ,
\label{pee-bulk}
\end{eqnarray}
where $h\nu = E$ is assumed.
\begin{table*}
 \centering
 \begin{minipage}{115mm}
  \caption{Physical parameters of grain materials;
The bulk work function ${W}_{\infty}$, the dielectric constant of the grain material $\epsilon$, the fitting constants for the mean free path of inelastic electron scattering $(a, b)$, the mean free path of inelastic electron scattering ${l}_{\mathrm{e}}$ at $E = 10.2~\mathrm{eV}$, and the photon attenuation length ${l}_{\mathrm{a}}$ at $h \nu = 10.2~\mathrm{eV}$. \label{physical-parameter}}
  \begin{tabular}{@{}lccccccl@{}}
  \hline
  & \multicolumn{1}{c}{${W}_{\infty}$} & \multicolumn{1}{c}{$\epsilon$} & \multicolumn{1}{c}{$a$} & \multicolumn{1}{c}{$b$} &  \multicolumn{1}{c}{${l}_{\mathrm{e}}(10.2~\mathrm{eV})$} & \multicolumn{1}{c}{${l}_{\mathrm{a}}(10.2~\mathrm{eV})$} &   \\
Material & \multicolumn{1}{c}{(eV)} & \multicolumn{1}{c}{(nm)} & \multicolumn{1}{c}{(nm)} & \multicolumn{1}{c}{(nm)} & \multicolumn{1}{c}{(nm)} & \multicolumn{1}{c}{(nm)} & Reference \\
  \hline
Silicate & 4.97 &  7.27 & 641 & 0.096 & 6.5 & 13 & 1, 2, 3, 4 \\
Silicon carbide & 7.0 & 9.72 & 641 & 0.096 & 6.5 & 5.0 & 1, 4, 5, 6 \\
Carbon & 4.75  & 7.0 & 143 & 0.054 & 1.5 & 15 & 1, 7, 8, 9 \\
Water ice & 8.7 & 107 & 641 & 0.096 & 6.5 & 40 & 1, 10, 11, 12 \\
Iron & 4.1  & $\infty$ & 143 & 0.054 & 1.5 & 9.2 & 1, 13, 14 \\
Organics & 5.75 &  6.9 & 31 & 0.087 & 0.58 & 15 & 1, 8, 15, 16 \\
\hline
\end{tabular}
\begin{flushleft}
References --- (1) \citet{seah-dench1979}; (2) \citet{feuerbacher-et-al1972}; (3) \citet{shannon-et-al1991}; (4) \citet{laor-draine1993}; (5) \citet{philipp1958}; (6) \citet{patrick-choyke1970}; (7) \citet{feuerbacher-fitton1972}; (8) \citet{louh-et-al2005}; (9) \citet{rouleau-martin1991}; (10) \citet{baron-et-al1978}; (11) \citet{johari-whalley1981}; (12) \citet{warren1984}; (13) \citet{heras-albano1983}; (14) \citet{moravec-et-al1976}; (15) \citet{fujihira-et-al1973}; (16) \citet{li-greenberg1997}.
\end{flushleft}
\end{minipage}
\end{table*}

\subsection{Material-dependent parameters \label{material}}

\begin{figure*}
\includegraphics[scale=0.8]{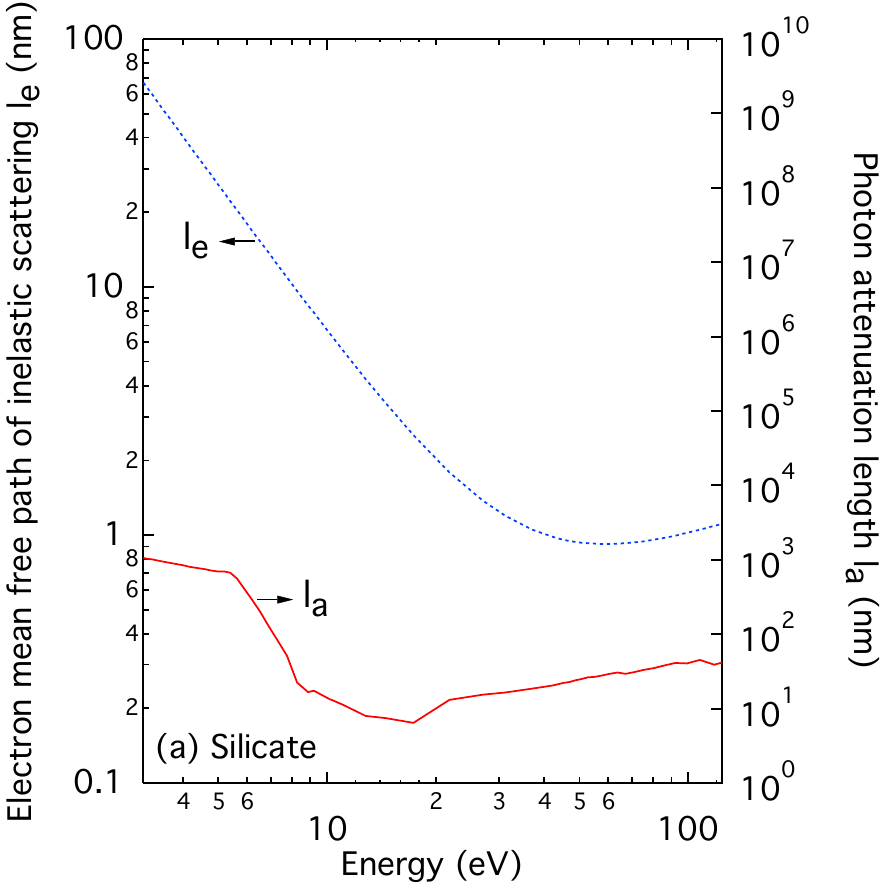}\includegraphics[scale=0.8]{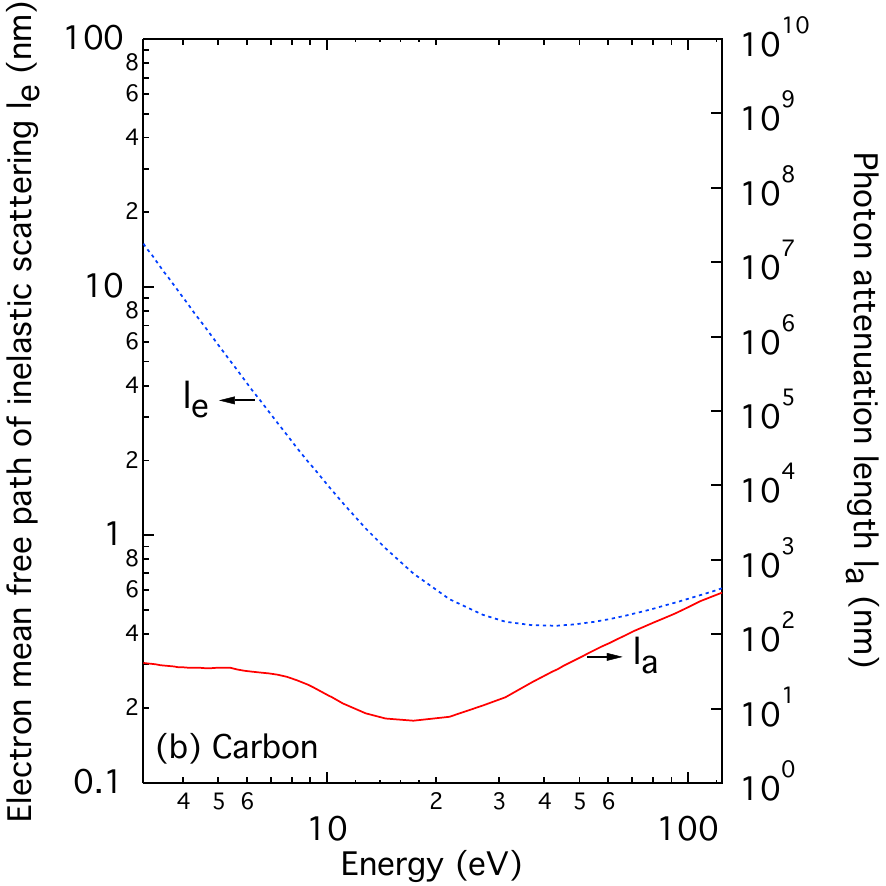}\\
\includegraphics[scale=0.8]{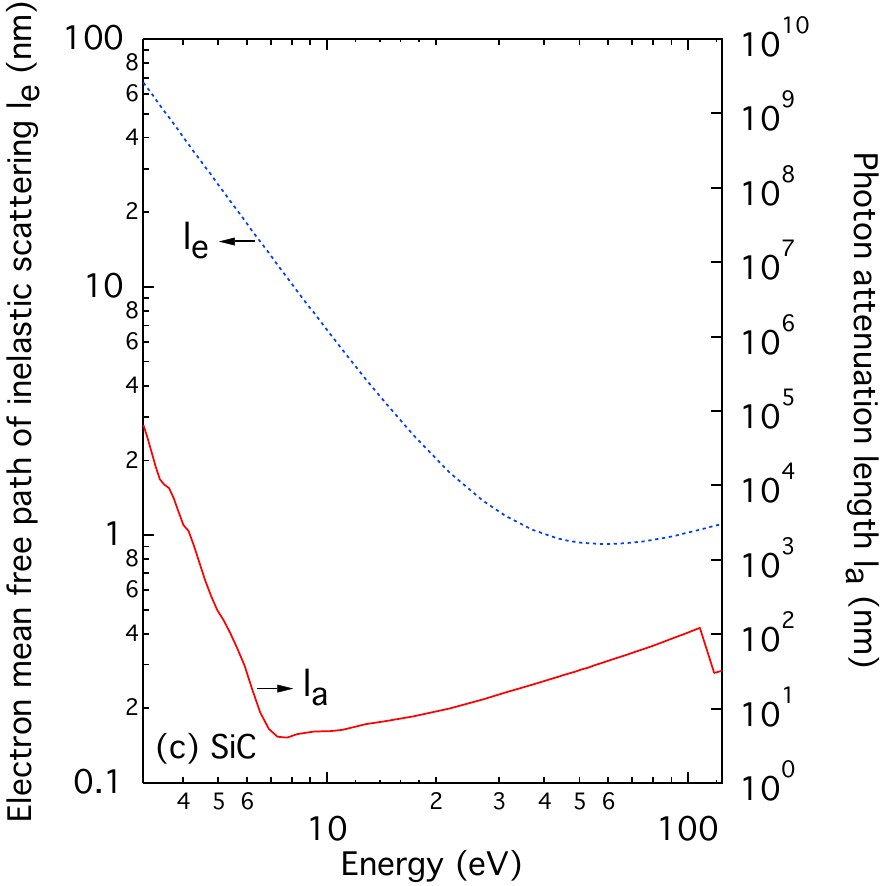}\includegraphics[scale=0.8]{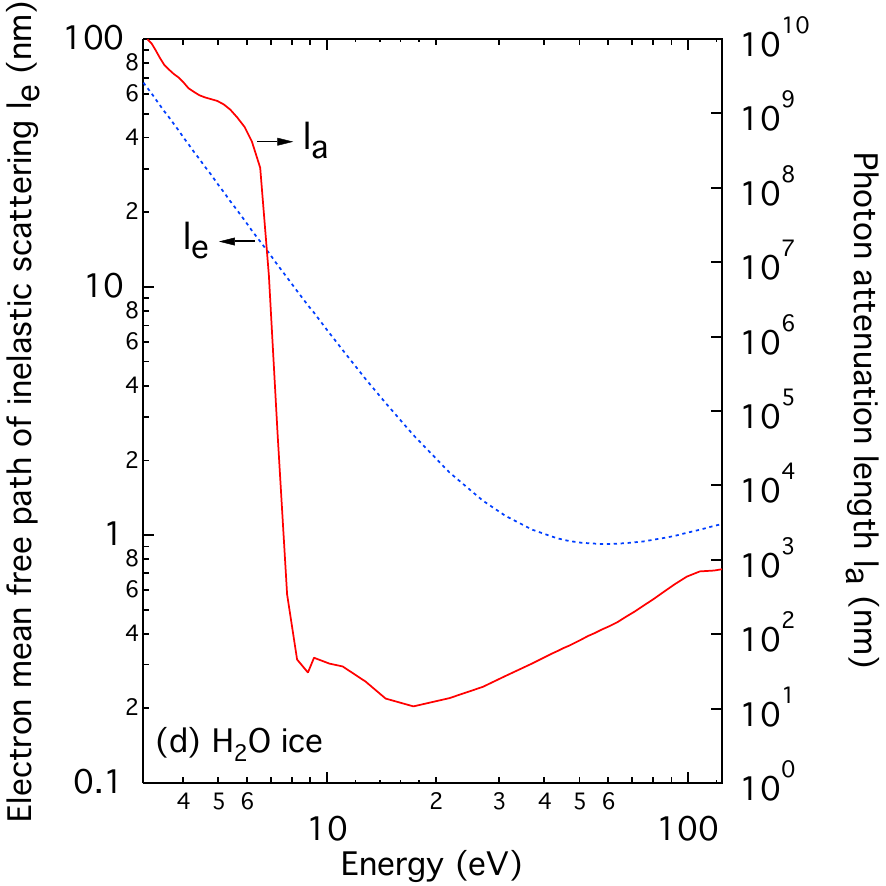}\\
\includegraphics[scale=0.8]{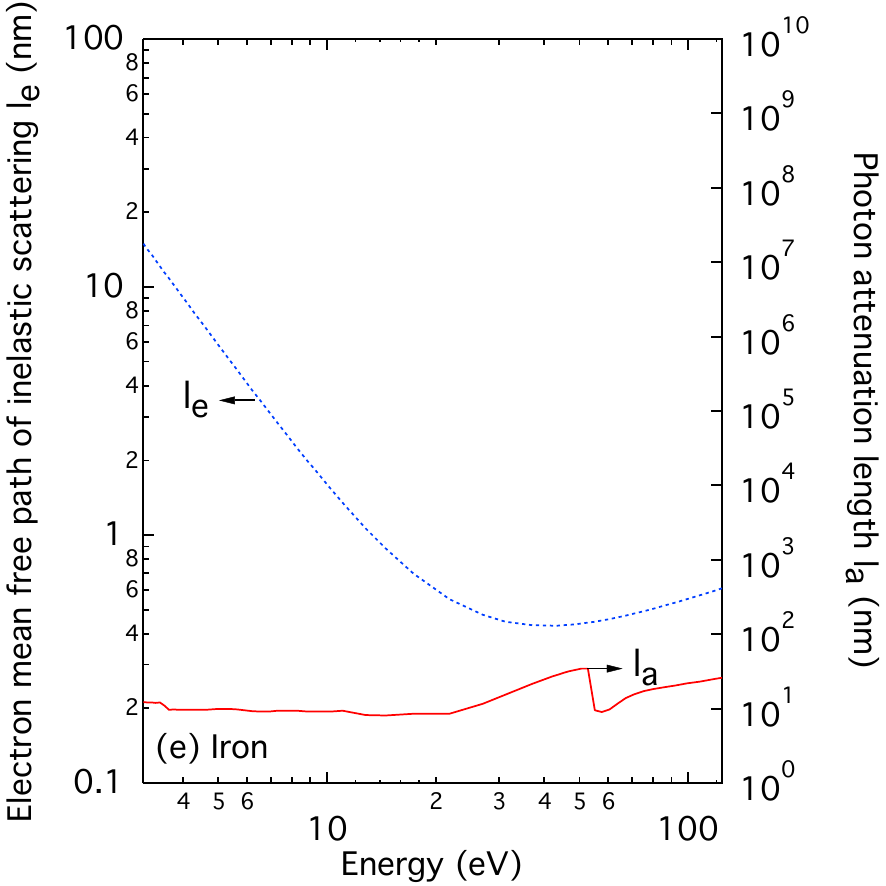}\includegraphics[scale=0.8]{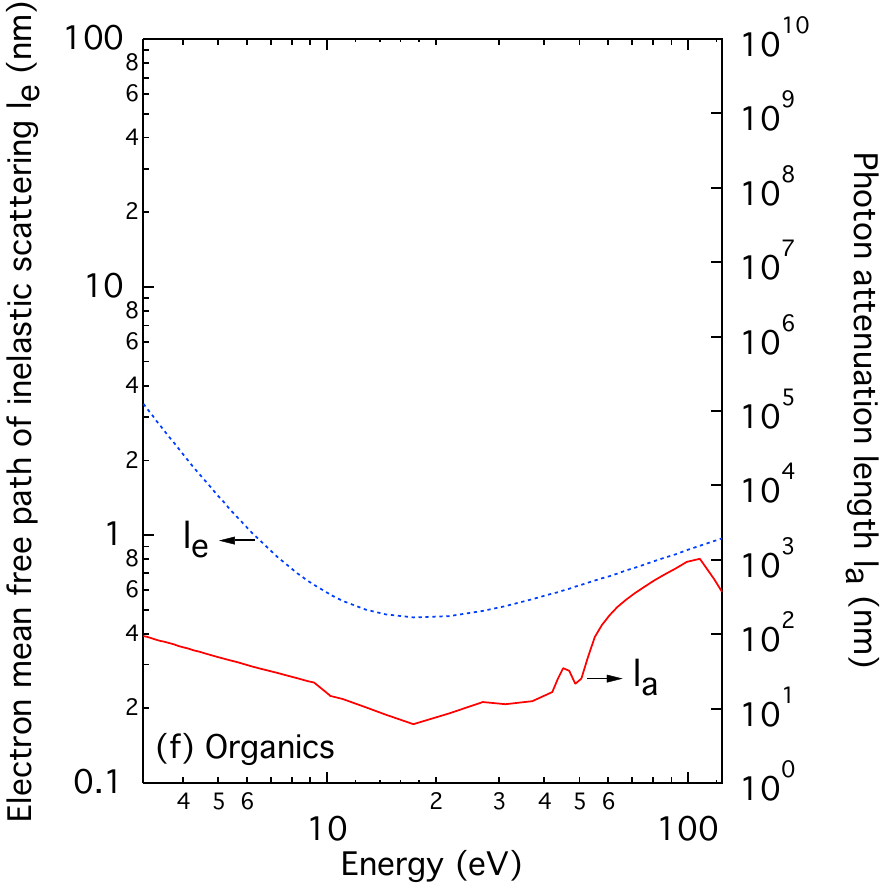}
\caption{The electron mean-free path of inelastic scattering ${l}_{\mathrm{e}}$ (dotted line) as a function of electron energy and the photon attenuation length ${l}_{\mathrm{a}}$ (solid line) as a function of photon energy for (a) silicate, (b) carbon, (c) silicon carbide, (d) water ice, (e) iron, and (f) organic material. \label{fig1}}
\end{figure*}

The complex refractive indices $m$, the work function for a bulk ${W}_{\infty}$, the static dielectric constant relative to vacuum $\epsilon$, and the electron mean-free path of inelastic scattering ${l}_{\mathrm{e}}$ are the parameters that depend on the composition of dust particles.
Let the particles be composed of silicate, silicon carbide, carbon, water ice, iron, or organics so that we could study the influence of dust materials on photoelectron emission.
The complex refractive indices $m$ of ``astronomical'' silicate and silicon carbide are taken from \citet{laor-draine1993}, those of amorphous carbon ``BE1'' from \citet{rouleau-martin1991}, and those of organic refractory material from \citet{li-greenberg1997}.
We also utilize the complex refractive indices for crystalline water ice from \citet{warren1984} in the range of photon energy $h \nu \le 28~\mathrm{eV}$ and extrapolate the indices to higher photon energies.
We combine the complex refractive indices of iron from \citet{moravec-et-al1976} in the range of photon energy $h \nu \le 27~\mathrm{eV}$ with those from \citet{palik1991} in the range of photon energy $h \nu > 27~\mathrm{eV}$.
The static dielectric constant relative to vacuum is given as the lowest frequency limit of relative permittivity: $\epsilon = 7.27$ for silicate, $\epsilon = 9.72$ for silicon carbide, $\epsilon = 7.0$ for carbon, $\epsilon = 107$ for water ice, $\epsilon = \infty$ for iron, and $\epsilon = 6.9$ for organics \citep{shannon-et-al1991,patrick-choyke1970,louh-et-al2005,johari-whalley1981}.
The work function for a semi-infinite slab has been derived from laboratory experiments: ${W}_{\infty} = 4.97~\mathrm{eV}$ for silicate, ${W}_{\infty} = 7.0~\mathrm{eV}$  for silicon carbide, ${W}_{\infty} = 4.75~\mathrm{eV}$ for carbon, ${W}_{\infty} = 8.7~\mathrm{eV}$ for water ice, ${W}_{\infty} = 4.1~\mathrm{eV}$ for iron, and ${W}_{\infty} = 5.75~\mathrm{eV}$ for organics \citep{feuerbacher-et-al1972,philipp1958,feuerbacher-fitton1972,baron-et-al1978,heras-albano1983,fujihira-et-al1973}.
The coefficients $a$ and $b$ in Eq.~(\ref{l_e}) for the inelastic mean-free path of electrons are $(a, b) = (641, 0.096)~\mathrm{nm}$ for silicate, silicon carbide, and water ice, $(a, b) = (31,0.087)~\mathrm{nm}$ for organics, and $(a, b) = (143, 0.054)~\mathrm{nm}$ for carbon and iron \citep{seah-dench1979}.
Table~\ref{physical-parameter} summarizes the physical parameters of these materials necessary for the calculations of photoelectric quantum yields.
The electron mean-free path of inelastic scattering ${l}_{\mathrm{e}}$ and the photon attenuation length ${l}_{\mathrm{a}}$ as a function of energy are plotted in Fig.~\ref{fig1} as dotted lines and solid lines, respectively.

\section{Results}

\begin{figure*}
\includegraphics[scale=0.8]{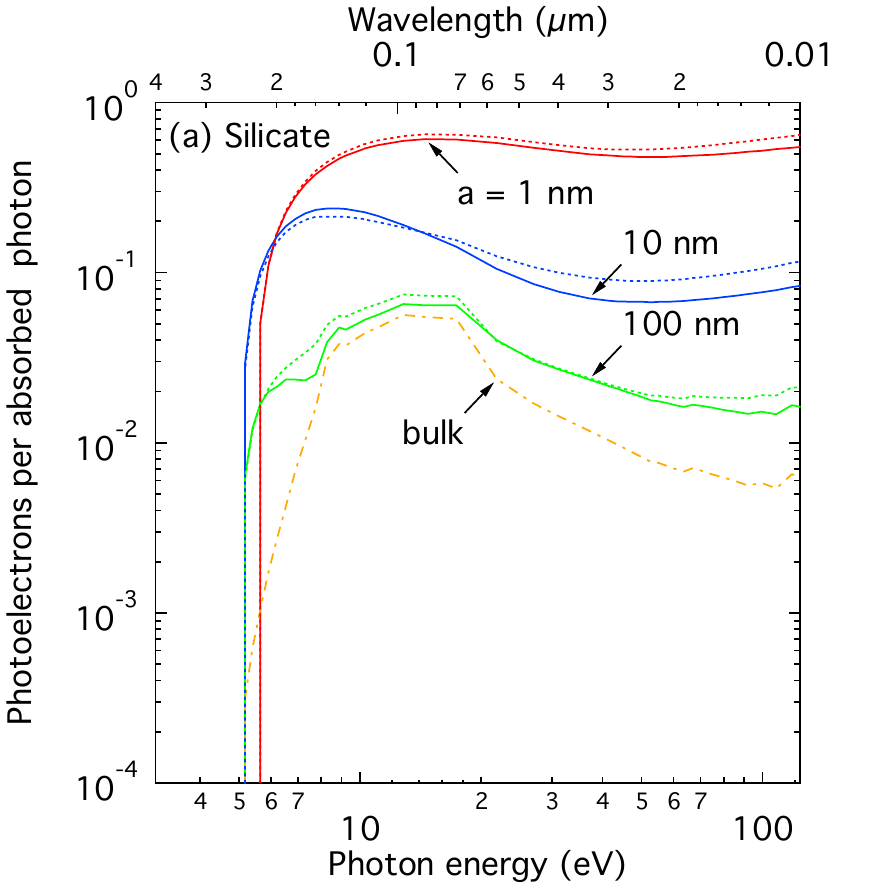}\includegraphics[scale=0.8]{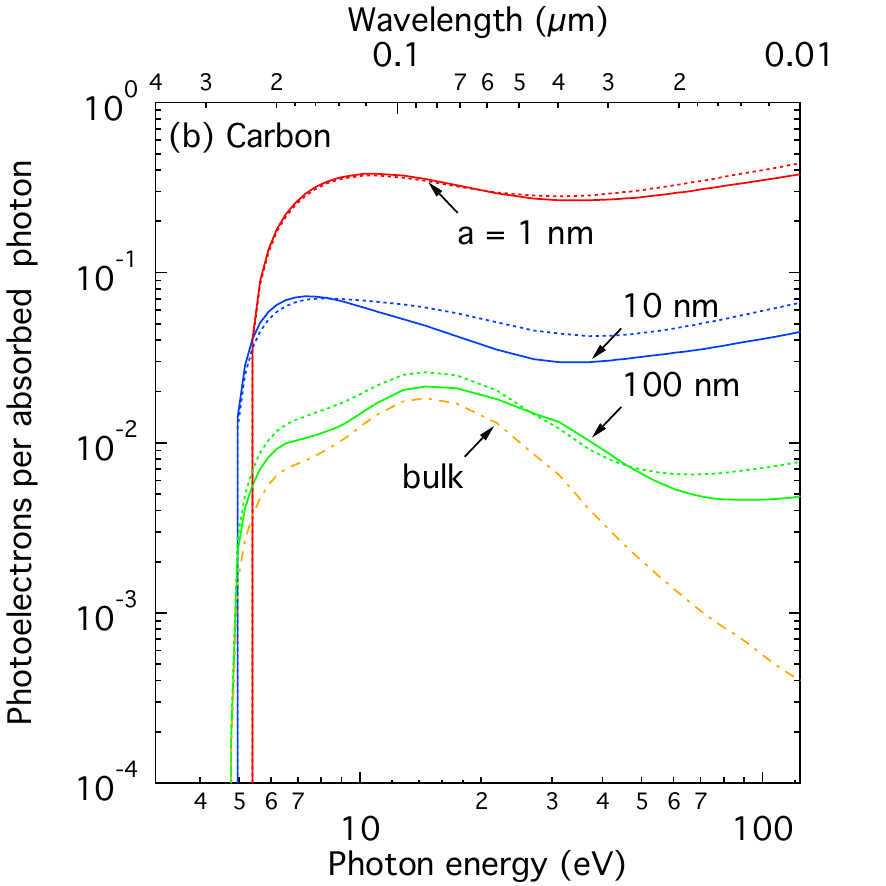}\\
\includegraphics[scale=0.8]{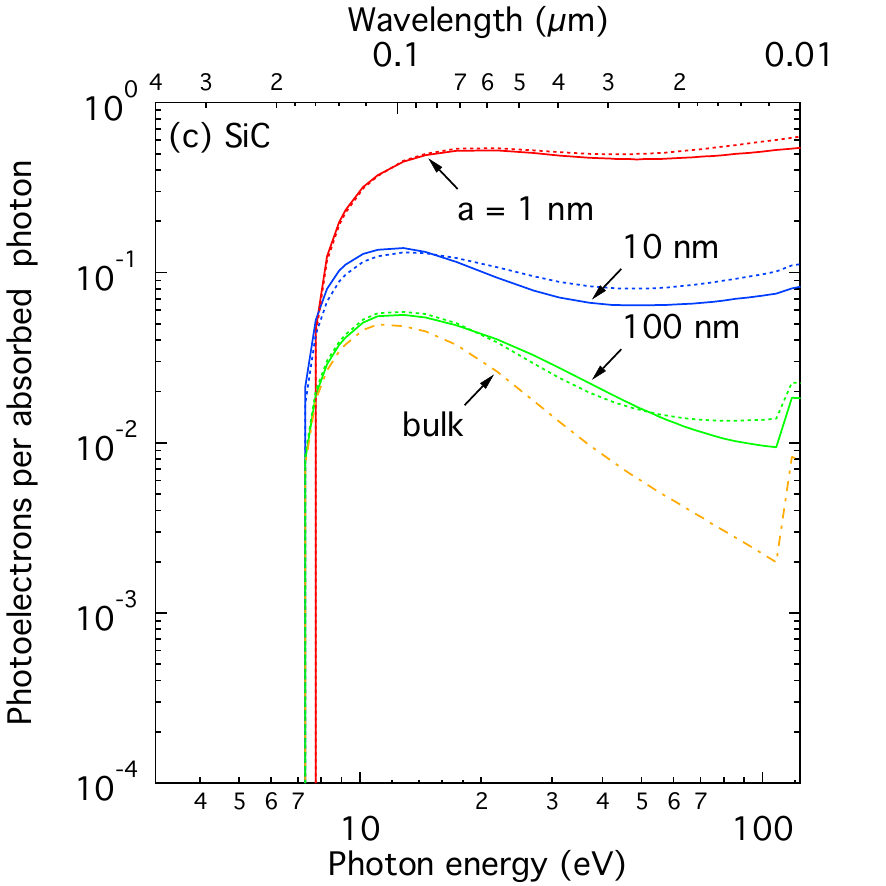}\includegraphics[scale=0.8]{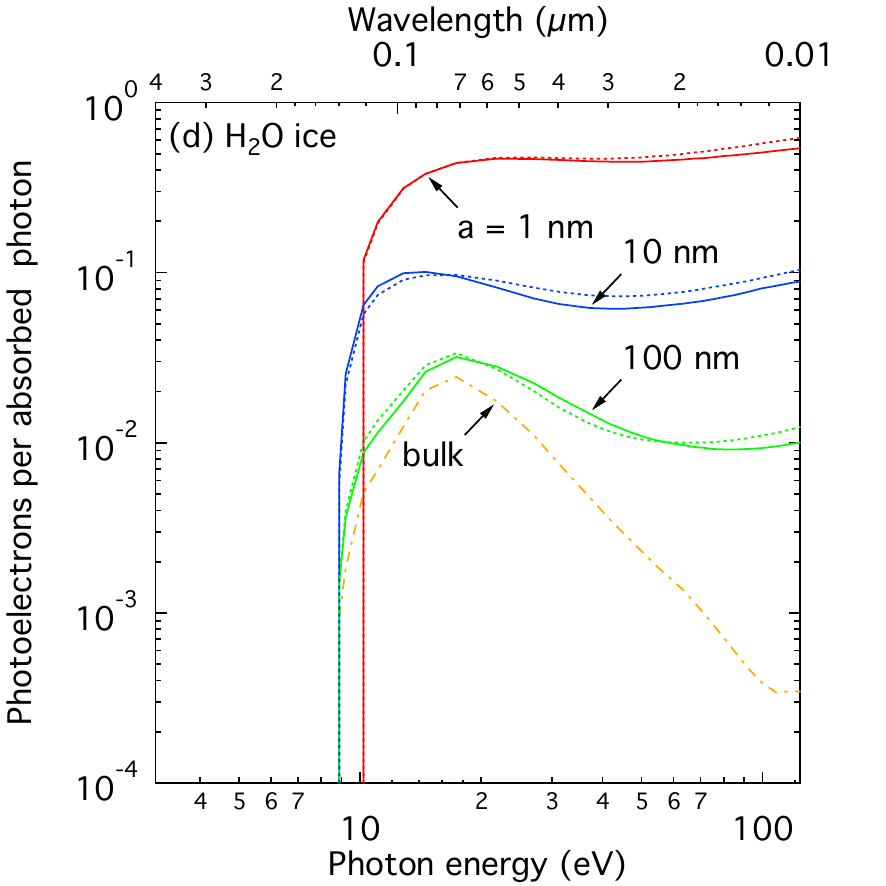}\\
\includegraphics[scale=0.8]{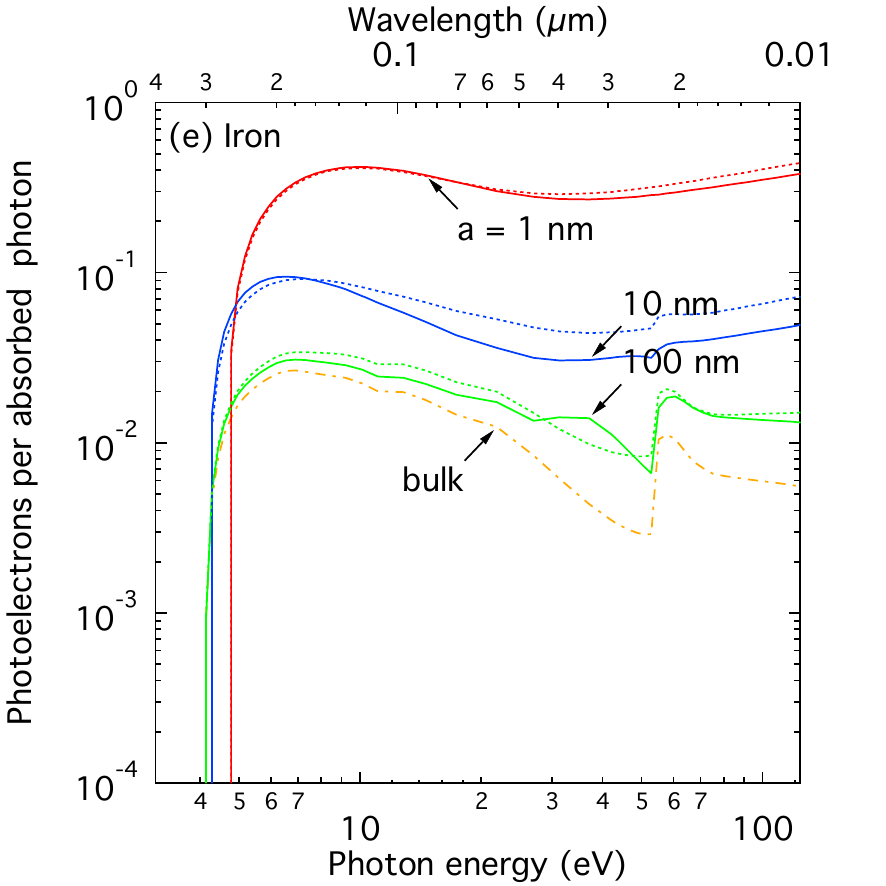}\includegraphics[scale=0.8]{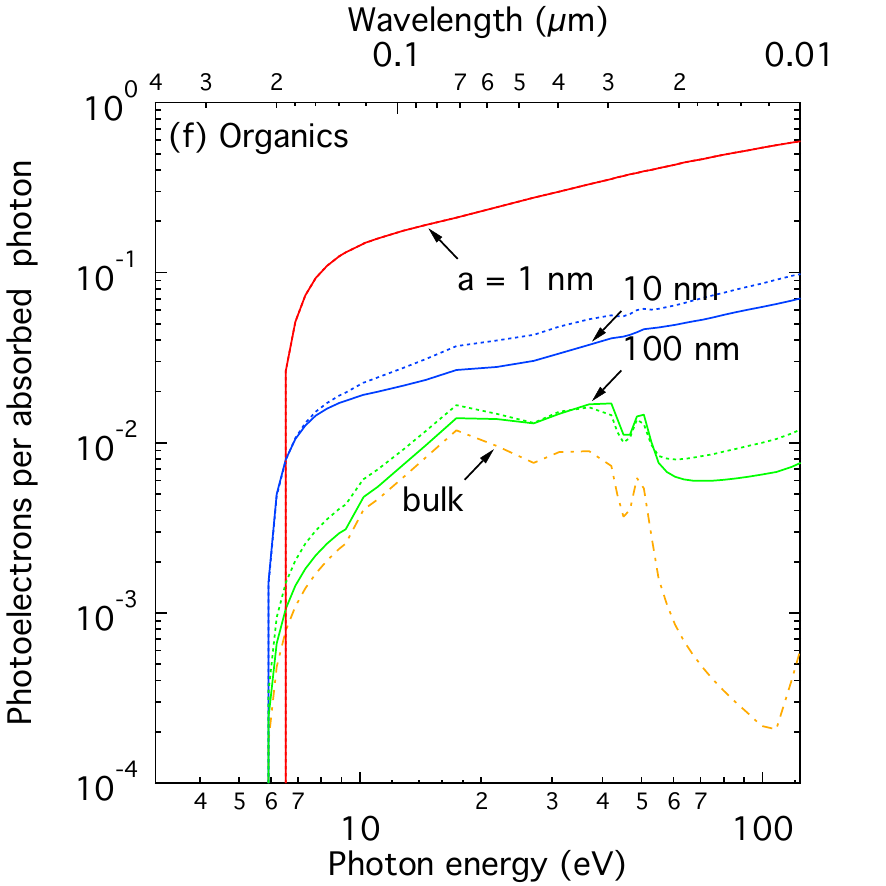}
\caption{Photoelectric quantum yields of dust particles composed of (a) silicate, (b) carbon, (c) silicon carbide, (d) water ice, (e) iron, and (f) organic material. Solid line: the exact formula of Eq.~(\ref{pee-sphere}) for spheres; dotted line: the approximate formula of Eq.~(\ref{our_formula}) for spheres; dash-dotted line: the exact formula of Eq.~(\ref{pee-bulk}) for bulk samples. \label{fig2}}
\end{figure*}
We calculate the photoelectric quantum yields of homogeneous spherical particles by performing numerical integration of Eq.~(\ref{pee-sphere}) whose numerical data are given in Tables~\ref{table-silicate}--\ref{table-organics}.
In Appendix~\ref{appendix-a}, we present an approximate formula of Eq.~(\ref{our_formula}) that reproduces the numerical data in Tables~\ref{table-silicate}--\ref{table-organics}.
Figure~\ref{fig2} shows our results on the photoelectric quantum yields of homogeneous spherical particles consisting of (a) silicate, (b) carbon, (c) silicon carbide, (d) water ice, (e) iron, and (f) organics as a function of photon energy.
Solid lines are the exact numerical integration of Eq.~(\ref{pee-sphere}), while dotted lines indicate the approximate formula of Eq.~(\ref{our_formula}).
Also shown as dash-dotted lines are the photoelectric quantum yields of semi-infinite slab surfaces calculated by the same three-step model given in Sect.~\ref{slab}.
Our results confirm that the small particle effect enhances the photoelectric quantum yields if the sizes of dust particles are smaller than submicrometers.
The photoelectric quantum yields of a bulk and a submicron grain show peaks at certain photon energies characteristic of the grain materials, originating from their refractive indices.
However, as the grain size decreases, such peaks disappear and the photoelectric quantum yields of nanoparticles become a smooth function of photon energy.
The photoelectric quantum yields of nanoparticles ($a=1~\mathrm{nm}$) have not yet reached asymptotic values even at a photon energy of $h\nu \simeq 100~\mathrm{eV}$, but become less dependent on grain materials at high energies of photons (i.e., $h\nu \gg W$).
For example, ${Y}_{a}(h\nu) \approx 0.5$ for silicate, silicon carbide, water ice, and organics, and ${Y}_{a}(h\nu) \approx 0.3$--0.4 for carbon and iron at a photon energy of $h\nu \simeq 100~\mathrm{eV}$.
\begin{table}
  \caption{The photoelectric quantum yields for silicate as a function of photon energy. Note that this table consists only of the first 5 rows of data; The full table is available online.\label{table-silicate}}
  \begin{tabular}{@{}cccccc@{}}
  \hline
   Energy     & Wavelength     & \multicolumn{4}{c}{Photoelectric quantum yield}\\
   (eV) & ($\umu$m) & \multicolumn{4}{c}{(e/absorbed)}  \\
     &   & $a=1\,$nm & $10\,$nm & $100\,$nm
     & $\infty$  \\
 \hline
4.95933  & 0.250000 & 0.00e+00 & 0.00e+00 & 0.00e+00 & 0.00e+00 \\
5.16597  & 0.240000 & 0.00e+00 & 2.94e-02 & 6.34e-03 & 3.15e-04 \\
5.39058  & 0.230000 & 0.00e+00 & 6.79e-02 & 1.21e-02 & 6.20e-04 \\
5.63561  & 0.220000 & 5.02e-02 & 1.03e-01 & 1.70e-02 & 1.04e-03 \\
5.90397  & 0.210000 & 1.11e-01 & 1.35e-01 & 2.00e-02 & 1.74e-03 \\
\vdots  & \vdots & \vdots & \vdots & \vdots & \vdots \\
\hline
\end{tabular}
\end{table}
\begin{table}
  \caption{The photoelectric quantum yields for carbon as a function of photon energy. Note that this table consists only of the first 5 rows of data; The full table is available online.\label{table-carbon}}
  \begin{tabular}{@{}cccccc@{}}
  \hline
   Energy     & Wavelength     & \multicolumn{4}{c}{Photoelectric quantum yield}\\
   (eV) & ($\umu$m) & \multicolumn{4}{c}{(e/absorbed)}  \\
     &   & $a=1\,$nm & $10\,$nm & $100\,$nm
     & $\infty$  \\
 \hline
4.59197  & 0.270000 & 0.00e+00 & 0.00e+00 & 0.00e+00 & 0.00e+00 \\
4.76859  & 0.260000 & 0.00e+00 & 0.00e+00 & 1.76e-04 & 1.50e-04 \\
4.95933  & 0.250000 & 0.00e+00 & 1.41e-02 & 2.31e-03 & 1.52e-03 \\
5.16597  & 0.240000 & 0.00e+00 & 2.85e-02 & 4.11e-03 & 2.68e-03 \\
5.39058  & 0.230000 & 4.06e-02 & 4.07e-02 & 5.63e-03 & 3.71e-03 \\
\vdots  & \vdots & \vdots & \vdots & \vdots & \vdots \\
\hline
\end{tabular}
\end{table}
\begin{table}
  \caption{The photoelectric quantum yields for silicon carbide as a function of photon energy. Note that this table consists only of the first 5 rows of data; The full table is available online.\label{table-sic}}
  \begin{tabular}{@{}cccccc@{}}
  \hline
   Energy     & Wavelength     & \multicolumn{4}{c}{Photoelectric quantum yield}\\
   (eV) & ($\umu$m) & \multicolumn{4}{c}{(e/absorbed)}  \\
     &   & $a=1\,$nm & $10\,$nm & $100\,$nm
     & $\infty$  \\
 \hline
6.88796  & 0.180000 & 0.00e+00 & 0.00e+00 & 0.00e+00 & 0.00e+00 \\
7.29314  & 0.170000 & 0.00e+00 & 2.15e-02 & 7.98e-03 & 7.54e-03 \\
7.74896  & 0.160000 & 4.63e-02 & 5.27e-02 & 1.91e-02 & 1.79e-02 \\
8.26556  & 0.150000 & 1.25e-01 & 8.05e-02 & 2.89e-02 & 2.67e-02 \\
8.85595  & 0.140000 & 1.99e-01 & 1.03e-01 & 3.77e-02 & 3.45e-02 \\
\vdots  & \vdots & \vdots & \vdots & \vdots & \vdots \\
\hline
\end{tabular}
\end{table}
\begin{table}
  \caption{The photoelectric quantum yields for water ice as a function of photon energy. Note that this table consists only of the first 5 rows of data; The full table is available online.\label{table-h2o}}
  \begin{tabular}{@{}cccccc@{}}
  \hline
   Energy     & Wavelength     & \multicolumn{4}{c}{Photoelectric quantum yield}\\
   (eV) & ($\umu$m) & \multicolumn{4}{c}{(e/absorbed)}  \\
     &   & $a=1\,$nm & $10\,$nm & $100\,$nm
     & $\infty$  \\
 \hline
8.26556  & 0.150000 & 0.00e+00 & 0.00e+00 & 0.00e+00 & 0.00e+00 \\
8.85595  & 0.140000 & 0.00e+00 & 6.60e-03 & 1.48e-03 & 9.49e-04 \\
9.18395  & 0.135000 & 0.00e+00 & 2.54e-02 & 3.66e-03 & 1.84e-03 \\
 10.1895 & 0.121677 & 1.18e-01 & 6.48e-02 & 8.77e-03 & 5.10e-03 \\
 10.1904 & 0.121667 & 1.19e-01 & 6.48e-02 & 8.77e-03 & 5.10e-03 \\
\vdots  & \vdots & \vdots & \vdots & \vdots & \vdots \\
\hline
\end{tabular}
\end{table}
\begin{table}
  \caption{The photoelectric quantum yields for iron as a function of photon energy. Note that this table consists only of the first 5 rows of data; The full table is available online.\label{table-iron}}
  \begin{tabular}{@{}cccccc@{}}
  \hline
   Energy     & Wavelength     & \multicolumn{4}{c}{Photoelectric quantum yield}\\
   (eV) & ($\umu$m) & \multicolumn{4}{c}{(e/absorbed)}  \\
     &   & $a=1\,$nm & $10\,$nm & $100\,$nm
     & $\infty$  \\
 \hline
3.99946  & 0.310000 & 0.00e+00 & 0.00e+00 & 0.00e+00 & 0.00e+00 \\
4.13278  & 0.300000 & 0.00e+00 & 0.00e+00 & 8.98e-04 & 9.28e-04 \\
4.27529  & 0.290000 & 0.00e+00 & 1.49e-02 & 5.22e-03 & 4.65e-03 \\
4.42798  & 0.280000 & 0.00e+00 & 3.08e-02 & 9.24e-03 & 8.11e-03 \\
4.59197  & 0.270000 & 0.00e+00 & 4.49e-02 & 1.29e-02 & 1.13e-02 \\
\vdots  & \vdots & \vdots & \vdots & \vdots & \vdots \\
\hline
\end{tabular}
\end{table}
\begin{table}
  \caption{The photoelectric quantum yields for organics as a function of photon energy. Note that this table consists only of the first 5 rows of data; The full table is available online.\label{table-organics}}
  \begin{tabular}{@{}cccccc@{}}
  \hline
   Energy     & Wavelength     & \multicolumn{4}{c}{Photoelectric quantum yield}\\
   (eV) & ($\umu$m) & \multicolumn{4}{c}{(e/absorbed)}  \\
     &   & $a=1\,$nm & $10\,$nm & $100\,$nm
     & $\infty$  \\
 \hline
5.63561  & 0.220000 & 0.00e+00 & 0.00e+00 & 0.00e+00 & 0.00e+00 \\
5.90397  & 0.210000 & 0.00e+00 & 1.51e-03 & 2.42e-04 & 1.76e-04 \\
6.19917  & 0.200000 & 0.00e+00 & 5.01e-03 & 6.63e-04 & 4.84e-04 \\
6.52544  & 0.190000 & 3.00e-02 & 8.00e-03 & 1.06e-03 & 7.89e-04 \\
6.88796  & 0.180000 & 5.78e-02 & 1.05e-02 & 1.45e-03 & 1.09e-03 \\
\vdots  & \vdots & \vdots & \vdots & \vdots & \vdots \\
\hline
\end{tabular}
\end{table}
\begin{figure*}
\includegraphics[scale=0.8]{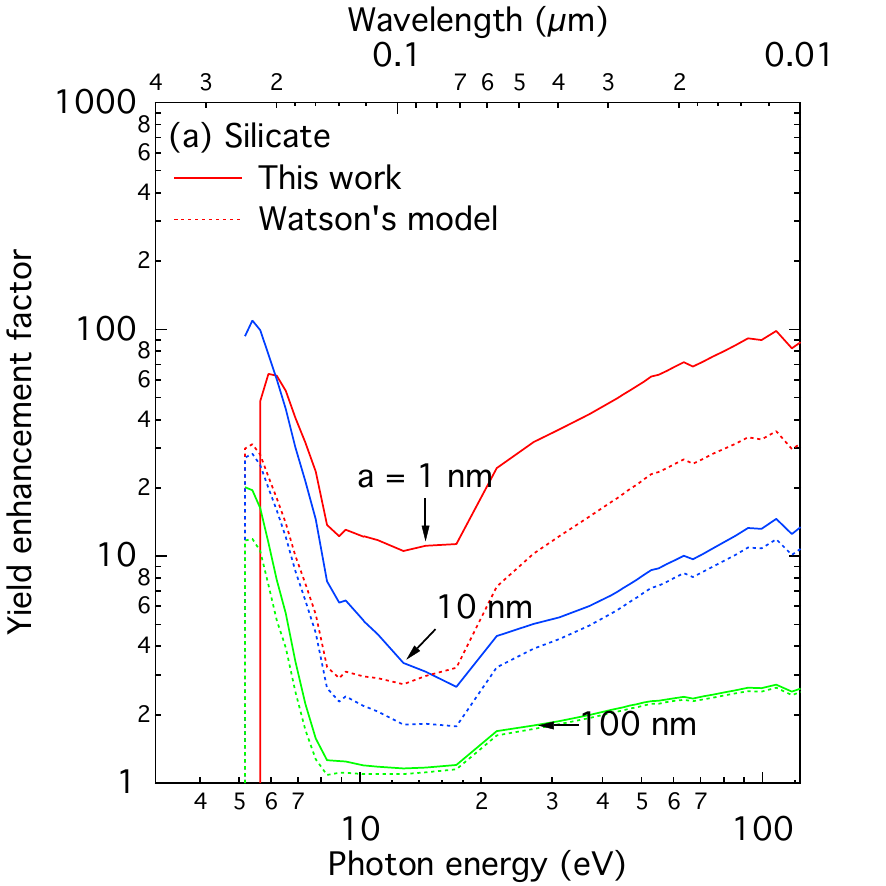}\includegraphics[scale=0.8]{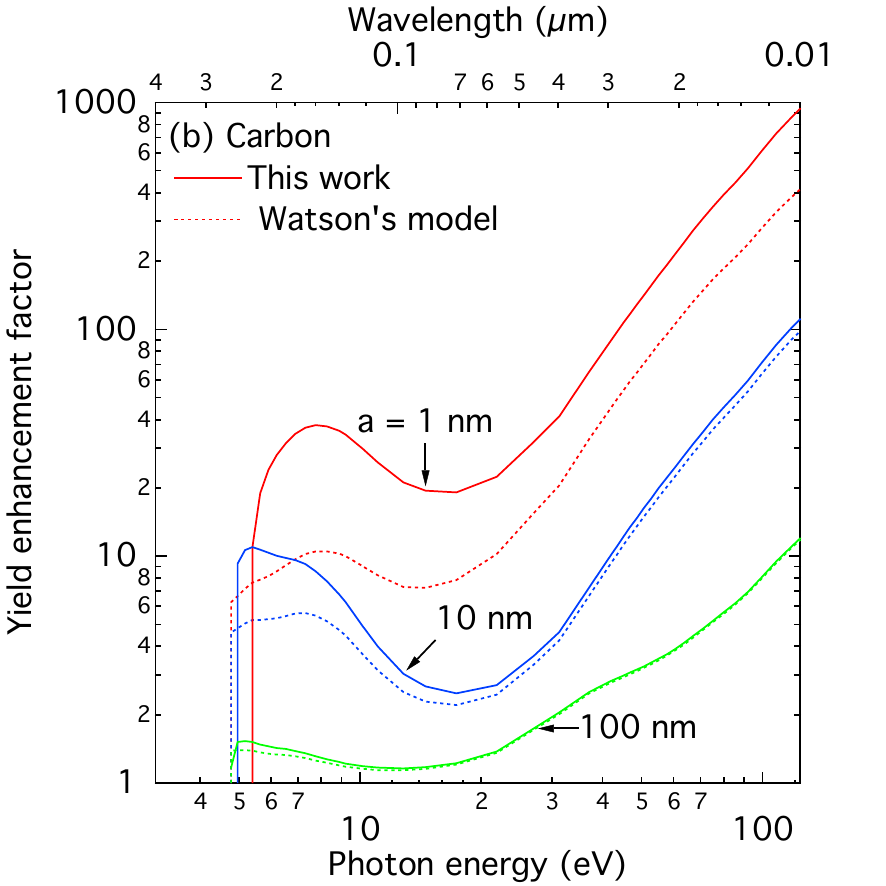}\\
\includegraphics[scale=0.8]{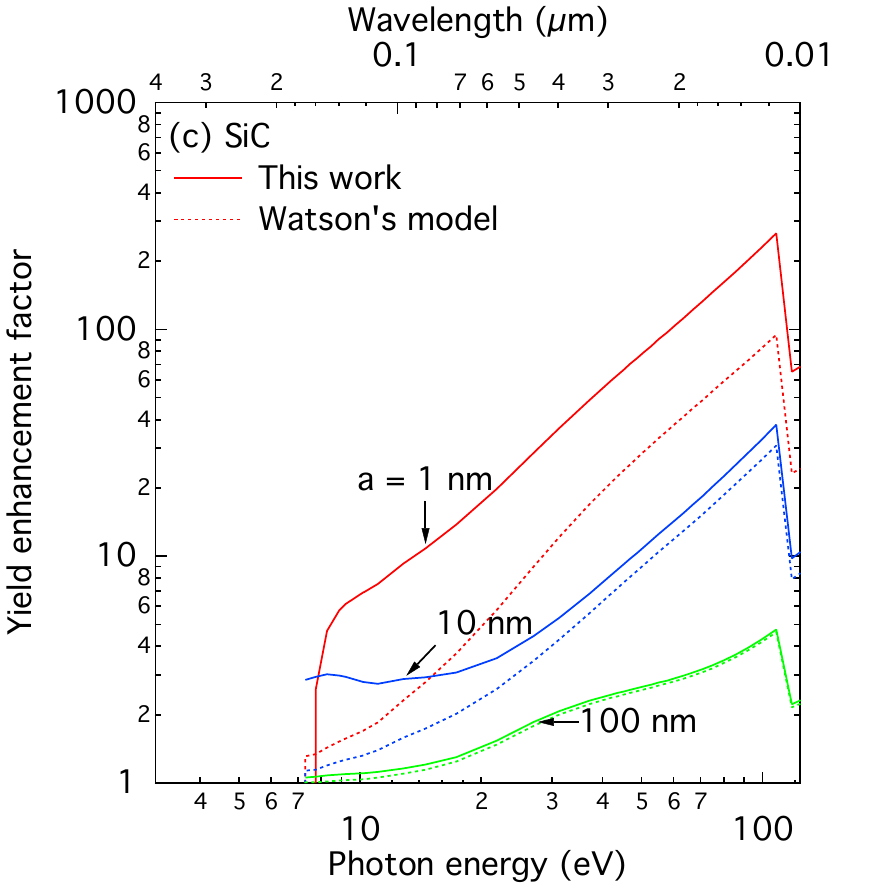}\includegraphics[scale=0.8]{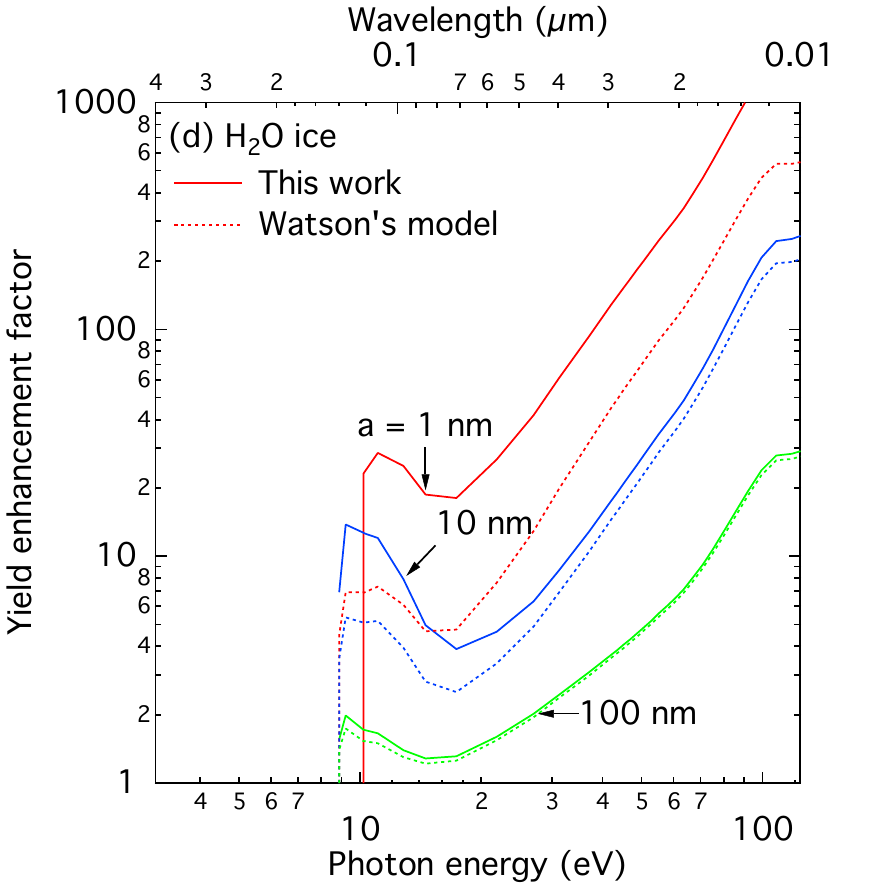}\\
\includegraphics[scale=0.8]{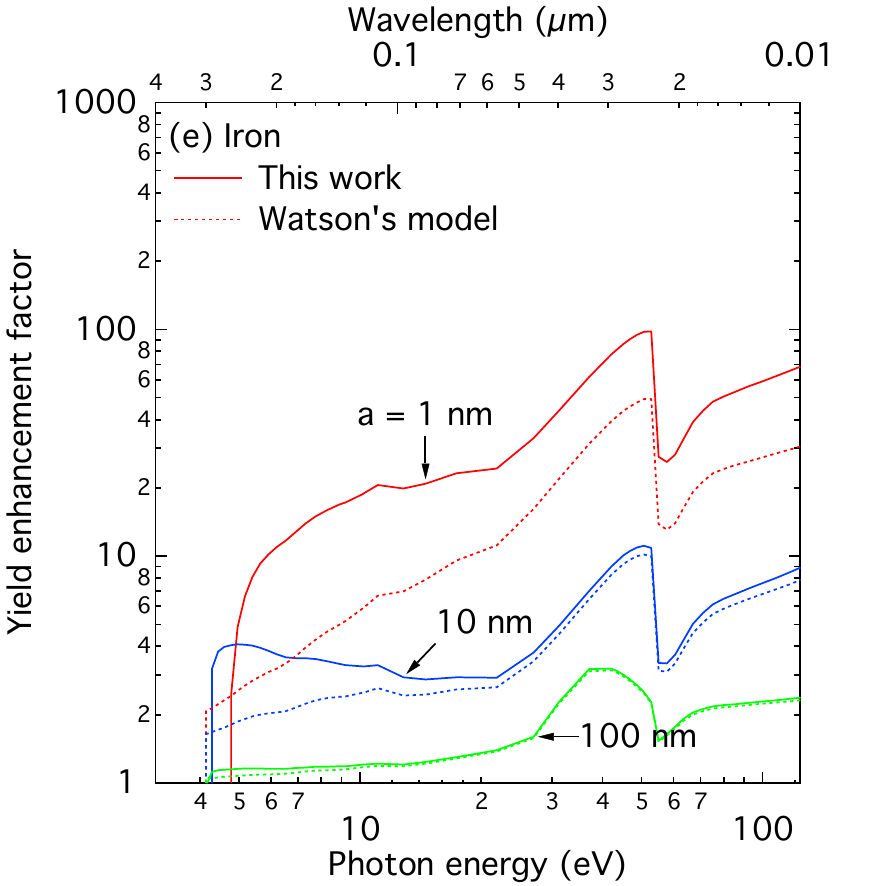}\includegraphics[scale=0.8]{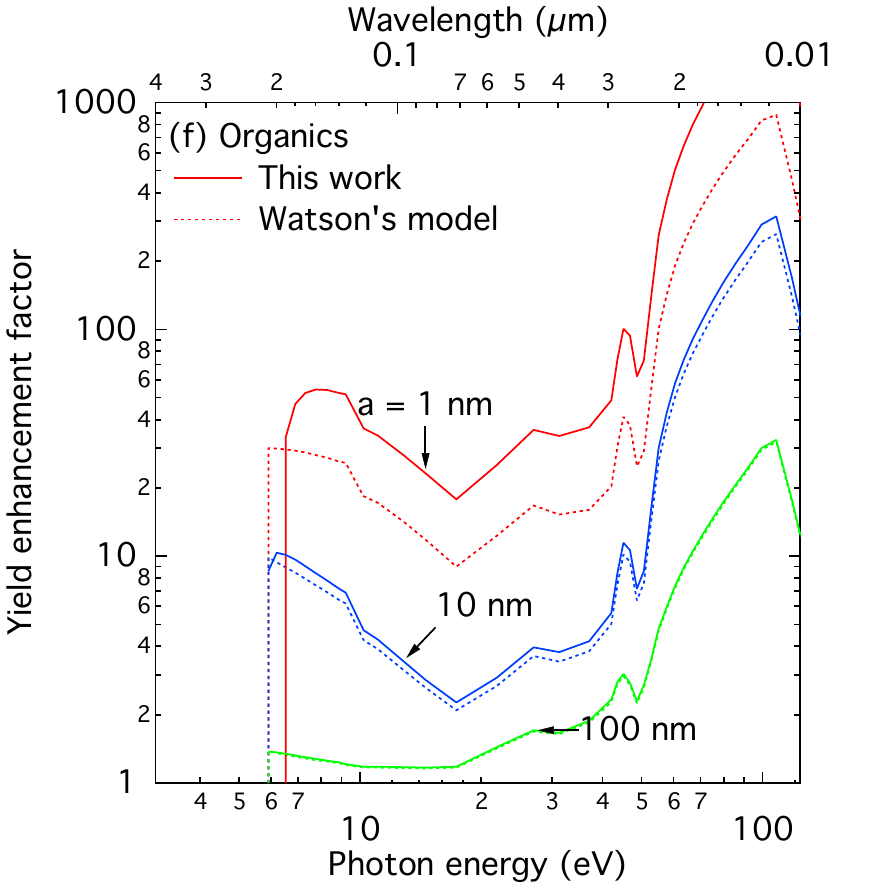}
\caption{Yield enhancement factors for dust particles composed of (a) silicate, (b) carbon, (c) silicon carbide, and (d) water ice, (e) iron, and (f) organic material. \label{fig3}}
\end{figure*}
Figure~\ref{fig3} illustrates the enhancement factors to the bulk photoelectric yields of homogeneous spherical particles consisting of (a) silicate, (b) carbon, (c) silicon carbide, (d) water ice, (e) iron, and (f) organics as a function of photon energy (solid lines). 
Also shown are the enhancement factors derived from \citeauthor{watson1973}'s model (dotted lines).
The difference in the enhancements between our model and \citeauthor{watson1973}'s model becomes noticeable as the radius of particles decreases.
Because \citeauthor{watson1973}'s model does not take into account the small particle effect of work function given in Eq.~(\ref{work-function}), the model also overestimates the photoelectric quantum yields near the thresholds.
The yield enhancement factors of small particles tend to increase toward high photon energies $h\nu > 15~\mathrm{eV}$, irrespective of dust compositions.
At high energies of $h\nu \ga 100~\mathrm{eV}$, the yield enhancement factors reach ${Y}_{a}(h\nu)/{Y }_{\infty}\left(h\nu \right) \sim 10^2$ for silicate, silicon carbide, and iron, but ${Y}_{a}(h\nu)/{Y }_{\infty}\left(h\nu \right) \ga 10^3$ for carbon, water ice, and organics.

\begin{figure*}
\includegraphics[scale=0.8]{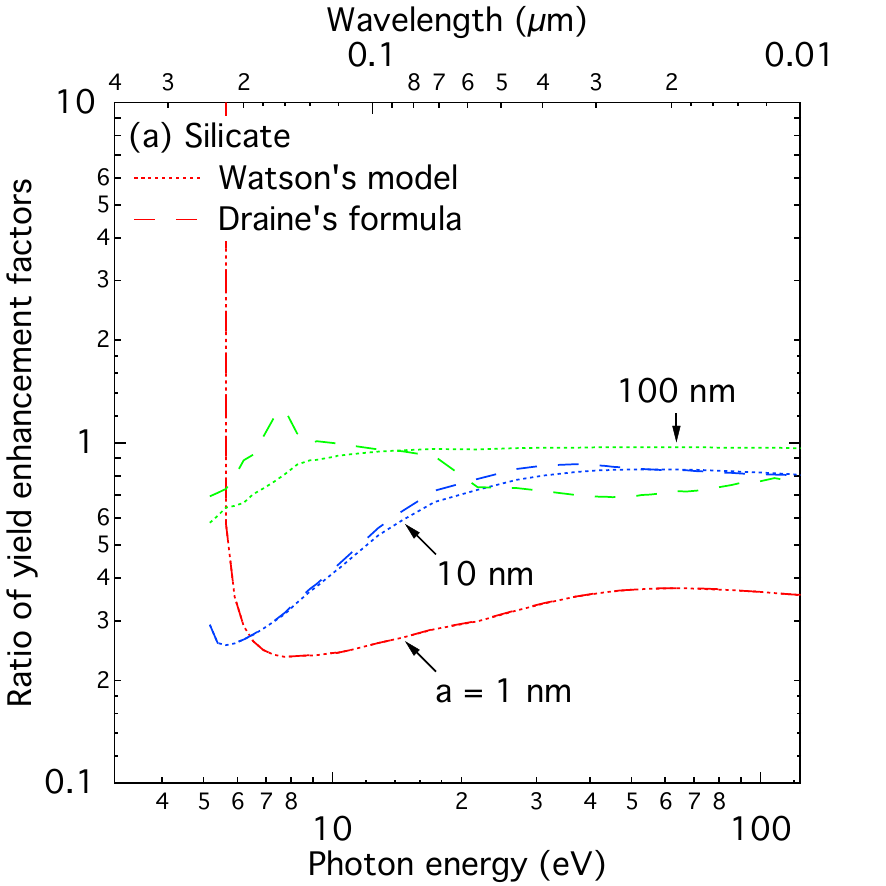}\includegraphics[scale=0.8]{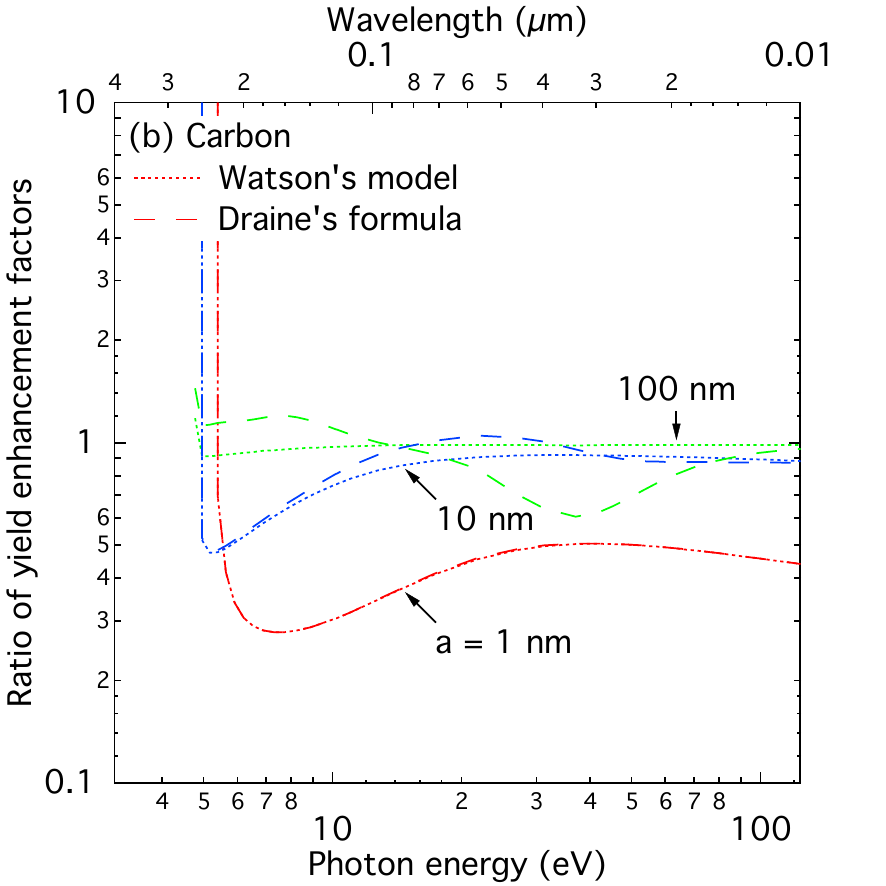}\\
\includegraphics[scale=0.8]{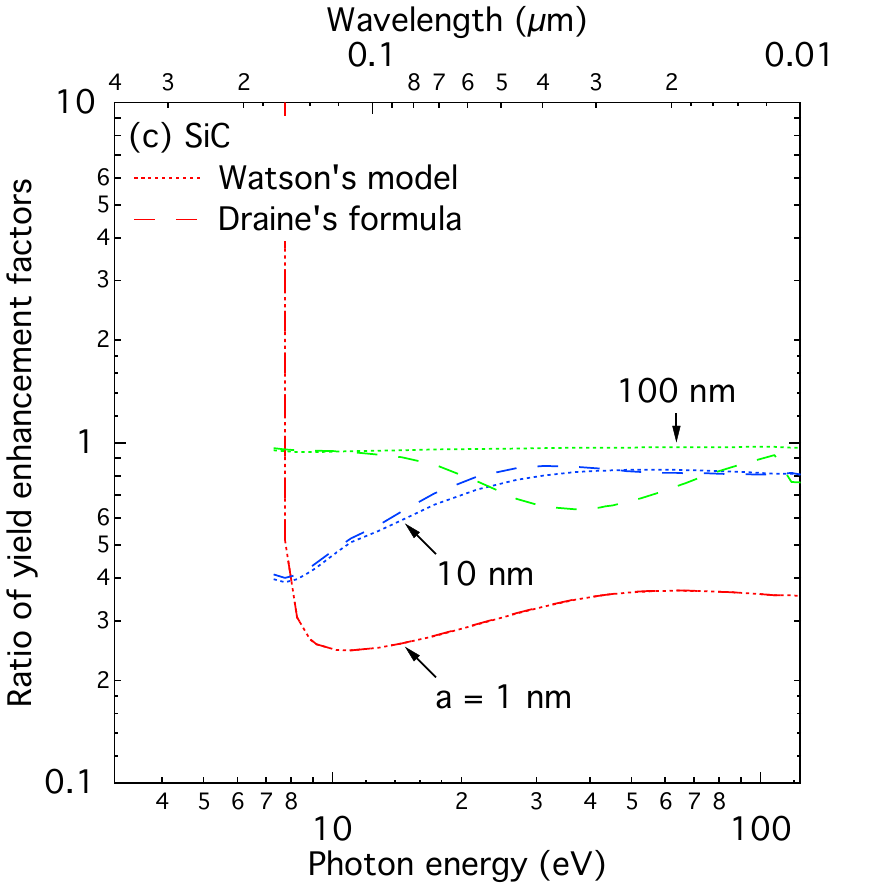}\includegraphics[scale=0.8]{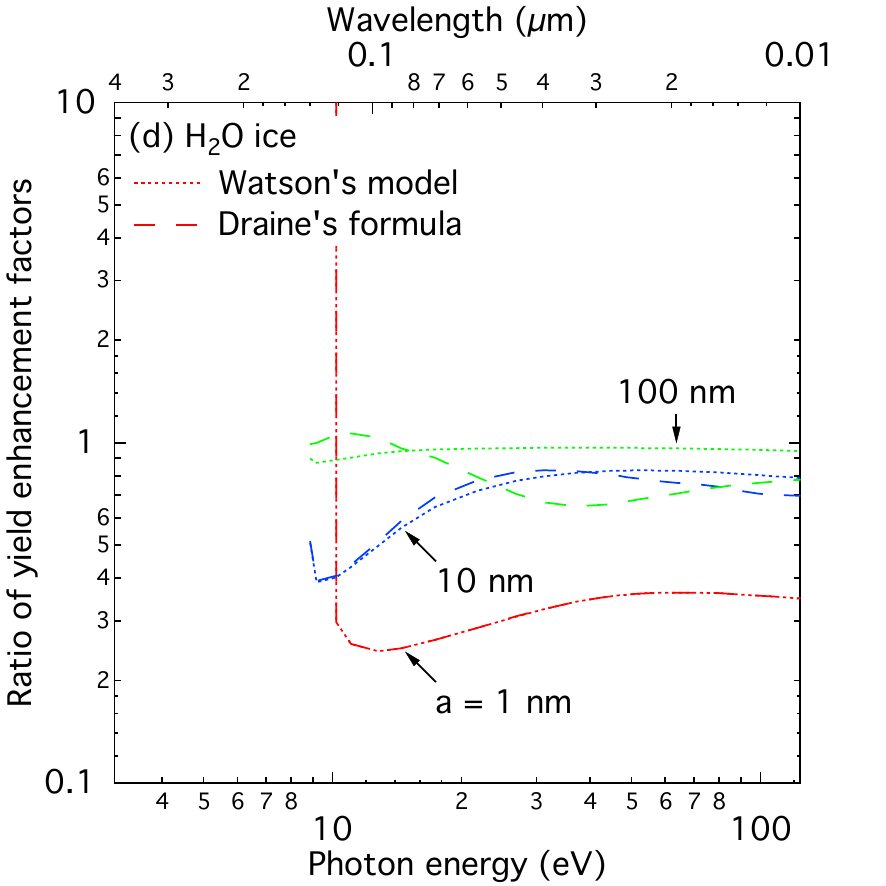}\\
\includegraphics[scale=0.8]{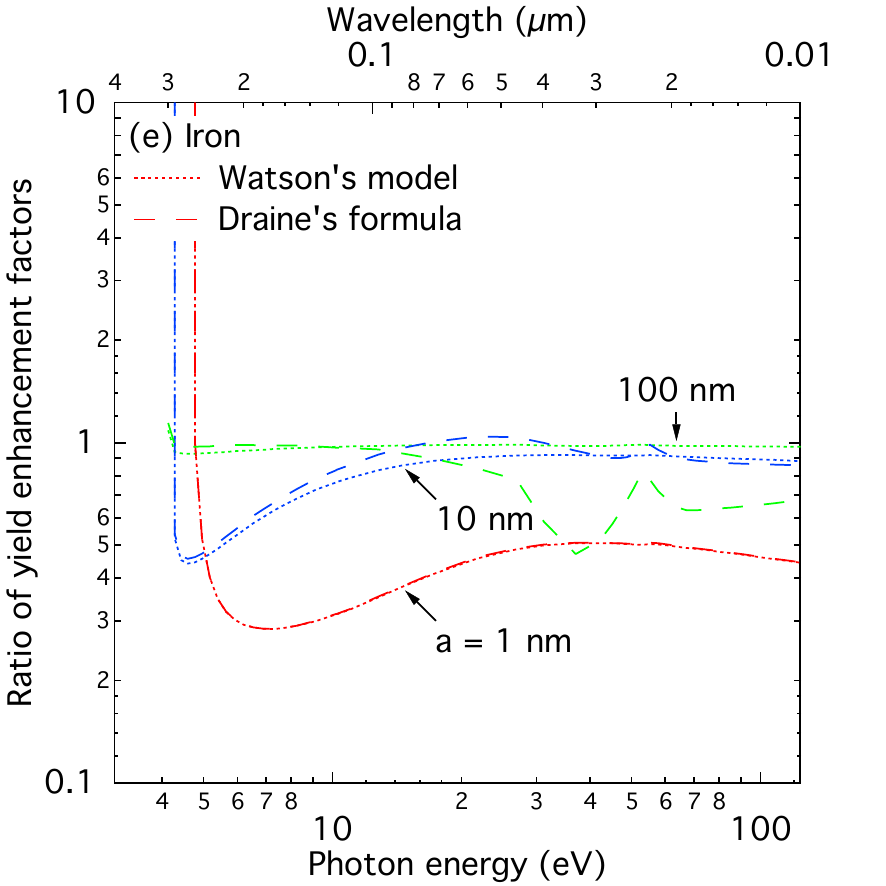}\includegraphics[scale=0.8]{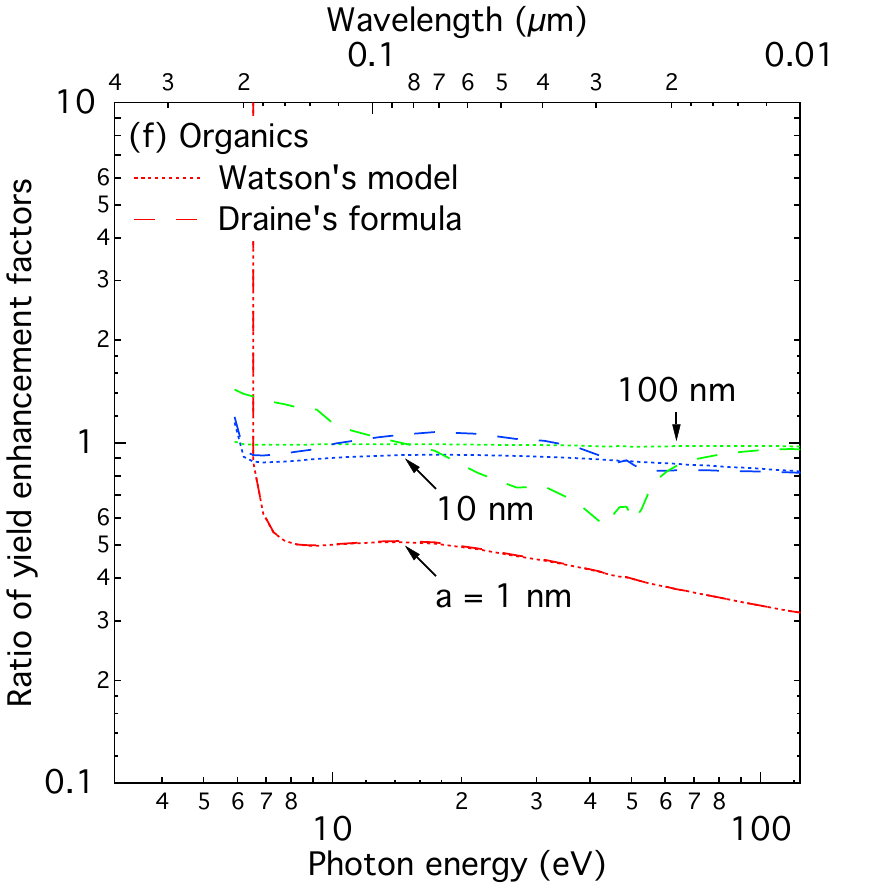}
\caption{Ratios of the yield enhancement factor derived from \citeauthor{watson1973}'s model (dotted line) or \citeauthor{draine1978}'s formula (dashed line) to that from our model for dust particles composed of (a) silicate, (b) carbon, (c) silicon carbide, and (d) water ice, (e) iron, and (f) organic material. \label{fig4}}
\end{figure*}
Figure~\ref{fig4} depicts how the deviations of \citeauthor{watson1973}'s model and \citeauthor{draine1978}'s formula in comparison to our model vary with photon energies and grain sizes.
The deviations are plotted as the ratios of the yield enhancement factor derived from \citeauthor{watson1973}'s model (dotted line) or \citeauthor{draine1978}'s formula (dashed line) to that from our model.
It turns our that \citeauthor{watson1973}'s model and \citeauthor{draine1978}'s formula result in lower values by a factor of 2--5 for the yield enhancement factors of nanoparticles ($a=1~\mathrm{nm}$).
For submicron grains ($a=100~\mathrm{nm}$), \citeauthor{watson1973}'s model reproduces our results to within a factor of 1.1 except for the energies close to the threshold (i.e., $h\nu \sim W$), while \citeauthor{draine1978}'s formula may deviate from our model values by a factor of 2 even at high photon energies (i.e., $h\nu \gg W$).
It is worthwhile mentioning that the deviations of Eq.~(\ref{our_formula}) from Eq.~(\ref{pee-sphere}) are kept below a factor of $1.2$ for nanoparticles irrespective of the energies and the materials, but may reach a factor of $1.6$ for submicron particles.

\section{Discussion}

We have taken into account the curvature of the particle surface for an estimate of the photoelectric quantum yield, which was ignored in \citeauthor{watson1973}'s model.
The curvature of the particle surface enhances the small particle effect of the photoelectric quantum yield because of an increase in the electron escape probability.
As a result, the difference between our model and \citeauthor{watson1973}'s model becomes significant as the radius of the particles decreases.
Although the difference between our model and \citeauthor{watson1973}'s model decreases with the radius of the particles, the deviations are noticeable even for $a=10~\mathrm{nm}$ at low photon energies.
While low-energy electrons generated by low-energy photons tend to suffer from the potential barrier at the surface, the curvature of the particle surface elevates the escape probability for electrons from the surface.
Consequently, the surface curvature plays a vital role in the determination of photoelectric quantum yields for not only small sizes, but also low-energy photons.

On the basis of \citeauthor{watson1973}'s model, \citet{ballester-et-al1995} estimated the yield enhancement factors ${Y}_{a}/{Y}_{\infty}$ of ``astronomical'' silicate grains and graphite grains with $a=10$ and $100~\mathrm{nm}$.
Subsequently, they determined the photoelectric quantum yields for the grains by multiplying the bulk yield ${Y}_{\infty}$ experimentally determined for lunar surface fines\footnote{Note that the absolute value of the photoelectric yield for lunar surface fines determined by \citet{feuerbacher-et-al1972} has been underestimated by a factor of two \citep[see][]{senshu-et-al2015}.} and graphite surface measured by \citet{feuerbacher-et-al1972} and \citet{feuerbacher-fitton1972}, respectively.
The numerical results of \citet{ballester-et-al1995} indicate that the photoelectric quantum yields of ``astronomical'' silicate grains and graphite grains lie in the range of ${10}^{-4}$ and ${10}^{-1}$ at $h\nu = 8$--$14~\mathrm{eV}$. 
In contrast, our results have revealed that the photoelectric quantum yields of ${Y}_{a}(h\nu) < {10}^{-3}$ appear only near the thresholds or much higher energies $h\nu > 50~\mathrm{eV}$. 
On the one hand, our model of the photoelectric quantum yields for small particles does not require the yield data for bulk samples, while the bulk yields determined in the framework of our model are self-consistent.
On the other hand, there is clearly an inconsistency in the complex refractive indices between ``astronomical'' silicate and lunar surface fines used in \citet{ballester-et-al1995}.
Therefore, it is ultimately important to determine the photoelectric quantum yields for small particles based on a self-consistent model.

A study on the dynamics of charged dust particles solves the equation of motion along with the evolution of electrical charge, which requires the knowledge of photoelectric current.
Because the photoelectric current is proportional to the photoelectric quantum yield, the small particle effect of photoelectron emission affects the determination of electrical charge.
However, the dynamics of charged nanoparticles has been studied commonly without consideration of the small particle effect of photoelectron emission \citep[e.g.,][]{Juhasz-Szego1998,mann-et-al2007,szego-et-al2014}.
We have shown that the small particle effect elevates photoelectric quantum yields so significantly that the effect cannot be neglected for nanoparticles (see Fig.~\ref{fig2}).
The omission of the surface curvature underestimates the photoelectric currents of nanoparticles by at least one order of magnitude as well as the grain charges to some extent.
If we calculate the electric charges on silicate nanoparticles with $a=1~\mathrm{nm}$ at 1~AU from the Sun as an example, then the equilibrium surface potential $U$ of the particles may reveal the importance of grain surface geometry \citep[see][for the computation of grain charges]{kimura-mann1998}.
The use of the photoelectric quantum yield modeled by \citet{draine-salpeter1979} results in $U=3.5~\mathrm{V}$, while it turns out that we obtain $U=3.2~\mathrm{V}$ using the bulk photoelectric yield of Eq.~(\ref{pee-bulk}) and the application of \citeauthor{watson1973}'s model (i.e., without the surface curvature) to the bulk yield results in $U=3.2~\mathrm{V}$\footnote{Note that \citeauthor{watson1973}'s model slightly elevates the surface potential, although the elevation is not obvious, owing to a crucial effect of secondary electron emission on the electric grain charging. Here the secondary electrons are produced by bombardments of high-energy electrons whose contribution to grain charging increases at an elevated surface potential.}.
In contrast, our model (i.e., with the surface curvature) gives $U=4.3~\mathrm{V}$, which clearly demonstrates the importance of the surface curvature.
Therefore, we claim that any forthcoming study on the dynamics of charged nanoparticles should implement the small particle effect with the surface curvature in the study.

A series of experimental experiments on the photoelectric quantum yields for submicron to micron-sized grains with astronomically relevant materials was performed by \citet{abbas-et-al2002,abbas-et-al2006,abbas-et-al2007} at photon energies of 7.8, 8.9, and $10.3~\mathrm{eV}$.
Their experimental results show that the photoelectric quantum yields increase with grain size, contrary to not only theoretical works but also previous experimental works on the small particle effect of photoelectron emission (cf. Fig.~\ref{fig2}).
\citet{abbas-et-al2006,abbas-et-al2007} claim that the photoelectric quantum yield of silver nanoparticles measured by \citet{mueller-et-al1988b} shows the size dependence consistent with their experiments.
However, we notice that the total number of photoelectrons ejected from a particle per {\sl incident} photon in the experiments of \citet{mueller-et-al1988b} does not show a clear difference between the particles with $a = 2.7$, 3.8, and $5.4~\mathrm{nm}$ within the uncertainties.
Note that the total number of photoelectrons ejected from a particle per {\sl incident} photon is given by ${Y}_{a}(h\nu)\, Q_\mathrm{abs}$ where $Q_\mathrm{abs}$ is the absorption efficiency of photons, because ${Y}_{a}(h\nu)$ is the total number of electrons ejected from a particle per {\sl absorbed} photon.
Because $Q_\mathrm{abs} \propto a$ for photon energies used their experiments, it is natural to expect that the photoelectric quantum yields of silver nanoparticles measured by \citet{mueller-et-al1988b} increase with decreasing grain size.
It is worthwhile mentioning that \citet{schmidtott-et-al1980} were the first to experimentally determine ${Y}_{a}(h\nu)\, Q_\mathrm{abs}$ near threshold for silver nanoparticles with $a=2.0$, 2.7, and $3.0~\mathrm{nm}$ and to show the increase in ${Y}_{a}(h\nu)\, Q_\mathrm{abs}$ with decreasing particle size.
Because of $Q_\mathrm{abs} \propto a$ for the nanoparticles with $a=2.0$, 2.7, and $3.0~\mathrm{nm}$ near the threshold, it is evident that the photoelectric quantum yields of silver nanoparticles decreases with particle size, at odds with the claim by \citet{abbas-et-al2006,abbas-et-al2007}.

From theoretical and past experimental points of view, the experiments of \citet{abbas-et-al2006,abbas-et-al2007} show contrary results on the size dependence of photoelectric quantum yields for small particles.
Furthermore, their asymptotic values of the photoelectric quantum yields for large grains are one to two orders of magnitude higher than those of bulk samples.
It is worthwhile noting that their estimates of photoelectric quantum yields require knowledge of discharge rates at zero grain charge, but they could measure the rates only down to about 10 electrons.
In the case of constant discharge rates, an extrapolation of the measured discharge rates to zero grain charge is remarkably straightforward.
However, this was not the case for the experiments of \citet{abbas-et-al2006,abbas-et-al2007}, although discharge rates for negatively charge particles under a fixed photon flux must be constant in theory \citep[cf.][]{mukai1981,kimura-mann1998,senshu-et-al2015}.
Therefore, we cannot help speculating that the discharge rates at zero grain charge have not been properly derived from their experiments, in particular, for micron-sized grains.

We have improved \citeauthor{watson1973}'s model for the small particle effect of photoelectron emission by taking the surface curvature of the particle into consideration, which has long been recognized as a missing piece of the model.
However, both the current study and \citeauthor{watson1973}'s model, which are based on the three step model, do not take into account the energy distribution of electrons at their production inside a particle.
Therefore, a more sophisticated model needs to include the electron energy distribution inside a particle, although the electron energy distribution must be known a priori.
While numerical demonstration of such a model is beyond the scope of this paper, here we briefly note the formula:
\begin{eqnarray}
{Y}_{a}(h\nu) = \frac{\int dV \, {\bf E}^{\ast}(r, \theta, h\nu) \cdot {\bf E}(r, \theta, h\nu) \, \int dE \, {p}_{\mathrm{esc}}(r, E) \,\eta(E, h\nu)}{\int dV \, {\bf E}^{\ast}(r, \theta, h\nu) \cdot {\bf E}(r, \theta, h\nu)} , 
\label{pee-sphere-correct}
\end{eqnarray}
where $\eta(E, h\nu) dE$ is the energy distribution of electrons in the energy range from $E$ to $E+dE$.
It is obvious that Eq.~(\ref{pee-sphere-correct}) reduces to Eq.~(\ref{pee-sphere}), when $\eta(E, h\nu) = \delta(E-h\nu)$ where $\delta$ is the Dirac $\delta$ function.
Consequently, if experimental data or theoretical models on $\eta(E, h\nu) dE$ for astronomically relevant materials are available, it is fairly straightforward to extend our model, which is the first step toward a more sophisticated model.

\section*{Acknowledgments}

I would like to thank Martin Hilchenbach, Harald Kr\"{u}ger, and Thomas Albin for their hospitality during my stay at Max Planck Institute for Solar System Research (MPS), where much of the writing was performed.
I am grateful to MPS's research fellowship and financial support for my travel expense to MPS as well as to JSPS's Grants-in-Aid for Scientific Research (\#26400230, \#15K05273, \#23103004).

\bsp

\label{lastpage}

\appendix

\section{Empirical analytic formula for photoelectric quantum yields of small particles}
\label{appendix-a}

\citet{draine1978} noted that \citeauthor{watson1973}'s numerical results are reproduced by the following analytic formula within 20\% of accuracy:
\begin{eqnarray}
\frac{{Y}_{a}(h\nu)}{{Y}_{\infty}(h\nu)} = \left({\frac{\beta}{\alpha}}\right)^{2}\frac{\alpha^2-2\alpha+2-2e^{-\alpha}}{\beta^2-2\beta+2-2e^{-\beta}},
\label{draine_formula}
\end{eqnarray}
where $\alpha = a/{l}_{\mathrm{a}} + a/{l}_{\mathrm{e}}$ and $\beta = a/{l}_{\mathrm{a}}$.
While the formula was presented without derivation, we notice that the formula can be derived from the analytic integration of the following approximation to \citeauthor{watson1973}'s formula:
\begin{eqnarray}
{Y}_{a}(h\nu) \approx \frac{\int_{0}^{a} {p}_{\mathrm{esc}}(r)\, \frac{1}{{l}_{\mathrm{a}}}\exp\left({-\frac{a-r}{{l}_{\mathrm{a}}}}\right) r^2 \, dr}{\int_{0}^{a}\frac{1}{{l}_{\mathrm{a}}}\exp\left({-\frac{a-r}{{l}_{\mathrm{a}}}}\right) r^2 \, dr},
\label{pee-sphere_approx}
\end{eqnarray}
where the escape probability ${p}_{\mathrm{esc}}(r)$ in the framework of \citeauthor{watson1973}'s model is given by
\begin{eqnarray}
{p}_{\mathrm{esc}}(r) = \frac{{l}_{\mathrm{a}}+{l}_{\mathrm{e}}}{{l}_{\mathrm{e}}} {Y}_{\infty}(h\nu) \exp\left({-\frac{a-r}{{l}_{\mathrm{e}}}}\right) .
\end{eqnarray}
By analogy, we may consider the escape probability ${p}_{\mathrm{esc}}(r)$ given by Eq.~(\ref{p_escape}) to take into account the surface curvature of small particles. 
Since the integration in Eq.~(\ref{p_escape}) cannot be performed analytically, we substitute Eq.~(\ref{l}) with $l \approx a-r$ to obtain
\begin{eqnarray}
{p}_{\mathrm{esc}}(r) \approx e^{-\left({\alpha-\beta}\right)} \left({1-\cos {\varphi }_{\mathrm{c}}}\right) \exp \left[{\left({\alpha-\beta}\right)\left({\frac{r}{a}}\right)}\right] .
\label{p_escape_approx}
\end{eqnarray}
Unfortunately, even if we substitute Eq.~(\ref{p_escape_approx}) into Eq.~(\ref{pee-sphere_approx}), the integration of the numerator in Eq.~(\ref{pee-sphere_approx}) cannot be performed analytically.
Consequently, we consider approximations for nanometer-sized particles ($a \la {l}_{\mathrm{e}}$) and submicrometer-sized particles ($a \ga {l}_{\mathrm{e}}$) separately, and then put them together into the final approximate formula.
For nanometer-sized particles ($a \la {l}_{\mathrm{e}}$), Eq.~(\ref{p_escape_approx}) is approximately
\begin{eqnarray}
{p}_{\mathrm{esc}}^{\mathrm{n}}(r) = \frac{1}{2} \left({1 - \frac{W}{E}}\right) e^{-\left({\alpha-\beta}\right)} {\left({\frac{r}{a}}\right)}^{-2} \exp \left[{\left({\alpha-\beta}\right)\left({\frac{r}{a}}\right)}\right] ,
\label{p_escape_small}
\end{eqnarray}
where the superscript $\mathrm{n}$ denotes the case for nanoparticles and we have replaced Eq.~(\ref{critical_angle}) by
\begin{eqnarray}
\cos {\varphi }_{\mathrm{c}} \approx 1 - \frac{1}{2} \left({1 - \frac{W}{E}}\right) {\left({\frac{r}{a}}\right)}^{-2} .
\label{critical_angle_small}
\end{eqnarray}
By substituting Eq.~(\ref{p_escape_small}) into Eq.~(\ref{pee-sphere_approx}), we obtain
\begin{eqnarray}
{Y}_{a}^{\mathrm{n}}(h\nu) = \frac{1}{2} \left({1-\frac{W}{h\nu}}\right) \left({\frac{\beta^{3} }{\alpha}}\right) \frac{1-e^{-\alpha}}{\beta^2 - 2\beta +2 - 2 e^{-\beta}}.
\label{yield-small}
\end{eqnarray}
For submicrometer-sized particles ($a \ga {l}_{\mathrm{e}}$), Eq.~(\ref{p_escape_approx}) is approximately
\begin{eqnarray}
{p}_{\mathrm{esc}}^{\mathrm{s}}(r) = \frac{1}{2} \left({1 - \sqrt{\frac{W}{E}}}\right) e^{-\left({\alpha-\beta}\right)} \exp \left[{\left({\alpha-\beta}\right)\left({\frac{r}{a}}\right)}\right] ,
\label{p_escape_large}
\end{eqnarray}
where the superscript $\mathrm{s}$ denotes the case for submicron particles and we consider an analogy to Eq.~(\ref{critical_angle_bulk}) for the critical angle:
\begin{eqnarray}
\cos {\varphi }_{\mathrm{c}} \approx \sqrt{\frac{W}{E}} .
\label{critical_angle_large}
\end{eqnarray}
The factor of $\frac{1}{2}$ originates from the fact that electrons could escape only from a hemisphere if the particles are sufficiently large compared to the mean free path of inelastic electron scattering.
It is worthwhile mentioning that $\frac{1}{2} \left({1 - \sqrt{W/E}}\right)$ is known as the semi-classical threshold function \citep{berglund-spicer1964a,smith1971}.
By substituting Eq.~(\ref{p_escape_large}) into Eq.~(\ref{pee-sphere_approx}), we obtain
\begin{eqnarray}
{Y}_{a}^{\mathrm{s}}(h\nu) = \frac{1}{2} \left({1 - \sqrt{\frac{W}{h\nu}}}\right)\left({\frac{\beta }{\alpha}}\right)^{3} \frac{\alpha^2 - 2\alpha + 2 - 2 e^{-\alpha}}{\beta^2 - 2\beta +2 - 2 e^{-\beta}} .
\label{yield-large}
\end{eqnarray}
To smoothly connect Eqs.~(\ref{yield-small}) and (\ref{yield-large}), we adopt the following equation:
\begin{eqnarray}
{Y}_{a}(h\nu) = \frac{1}{\alpha-\beta+1} \left[{{Y}_{a}^{\mathrm{n}}(h\nu) + {Y}_{a}^{\mathrm{s}}(h\nu) \left({\alpha-\beta}\right)}\right] ,
\end{eqnarray}
so that ${Y}_{a}(h\nu)$ reduces to ${Y}_{a}^{\mathrm{n}}(h\nu)$ for $a \ll {l}_{\mathrm{e}}$, while for $a \to \infty$, ${Y}_{a}(h\nu) \to {Y}_{a}^{\mathrm{s}}(h\nu)$.
Finally, we obtain an approximation to Eq.~(\ref{pee-sphere}) as follows:
\begin{eqnarray}
{Y}_{a}(h\nu) = \frac{1}{2} \left({1 - \sqrt{\frac{W}{h\nu}}}\right) \left({\frac{\beta }{\alpha}}\right)^{3} \frac{\left({1 + \sqrt{{W}/{h\nu}}}\right) \alpha^2 \left({1-e^{-\alpha}}\right) + \left({\alpha^2 - 2\alpha + 2 - 2 e^{-\alpha}}\right) \left({\alpha-\beta}\right)}{\left({\alpha-\beta+1}\right) \left({\beta^2 - 2\beta +2 - 2 e^{-\beta}}\right)} .
\label{our_formula}
\end{eqnarray}

\end{document}